\documentclass[12pt]{article}

\begin{document}

\def\CA{{\cal A}}
\def\CB{{\cal B}}
\def\CC{{\cal C}}
\def\CD{{\cal D}}
\def\CE{{\cal E}}
\def\CF{{\cal F}}
\def\CG{{\cal G}}
\def\CH{{\cal H}}
\def\CI{{\cal I}}
\def\CJ{{\cal J}}
\def\CK{{\cal K}}
\def\CL{{\cal L}}
\def\CM{{\cal M}}
\def\CN{{\cal N}}
\def\CO{{\cal O}}
\def\CP{{\cal P}}
\def\CQ{{\cal Q}}
\def\CR{{\cal R}}
\def\CS{{\cal S}}
\def\CT{{\cal T}}
\def\CU{{\cal U}}
\def\CV{{\cal V}}
\def\CW{{\cal W}}
\def\CX{{\cal X}}
\def\CY{{\cal Y}}
\def\CZ{{\cal Z}}

\newcommand{\todo}[1]{{\em \small {#1}}\marginpar{$\Longleftarrow$}}
\newcommand{\labell}[1]{\label{#1}}
\newcommand{\bbibitem}[1]{\bibitem{#1}}
\newcommand{\llabel}[1]{\label{#1}\marginpar{#1}}

\newcommand{\sphere}[0]{{\rm S}^3}
\newcommand{\su}[0]{{\rm SU(2)}}
\newcommand{\so}[0]{{\rm SO(4)}}
\newcommand{\bK}[0]{{\bf K}}
\newcommand{\bL}[0]{{\bf L}}
\newcommand{\bR}[0]{{\bf R}}
\newcommand{\tK}[0]{\tilde{K}}
\newcommand{\tL}[0]{\bar{L}}
\newcommand{\tR}[0]{\tilde{R}}

\newcommand{\btzm}[0]{BTZ$_{\rm M}$}
\newcommand{\ads}[1]{{\rm AdS}_{#1}}
\newcommand{\ds}[1]{{\rm dS}_{#1}}
\newcommand{\eds}[1]{{\rm EdS}_{#1}}
\newcommand{\sph}[1]{{\rm S}^{#1}}
\newcommand{\gn}[0]{G_N}
\newcommand{\SL}[0]{{\rm SL}(2,R)}
\newcommand{\cosm}[0]{R}
\newcommand{\hdim}[0]{\bar{h}}
\newcommand{\bw}[0]{\bar{w}}
\newcommand{\bz}[0]{\bar{z}}
\newcommand{\be}{\begin{equation}}
\newcommand{\ee}{\end{equation}}
\newcommand{\bea}{\begin{eqnarray}}
\newcommand{\eea}{\end{eqnarray}}
\newcommand{\pat}{\partial}
\newcommand{\lp}{\lambda_+}
\newcommand{\bx}{ {\bf x}}
\newcommand{\bk}{{\bf k}}
\newcommand{\bb}{{\bf b}}
\newcommand{\BB}{{\bf B}}
\newcommand{\tp}{\tilde{\phi}}
\hyphenation{Min-kow-ski}

\def\apr{\alpha'}
\def\str{{str}}
\def\lstr{\ell_\str}
\def\gstr{g_\str}
\def\Mstr{M_\str}
\def\lpl{\ell_{pl}}
\def\Mpl{M_{pl}}
\def\varep{\varepsilon}
\def\del{\nabla}
\def\grad{\nabla}
\def\tr{\hbox{tr}}
\def\perp{\bot}
\def\half{\frac{1}{2}}
\def\p{\partial}
\def\perp{\bot}
\def\eps{\epsilon}

\renewcommand{\thepage}{\arabic{page}}
\setcounter{page}{1}

\rightline{hep-th/0110108}
\rightline{VPI-IPPAP-01-01}
\rightline{UPR-964-T, 
ITFA-2001-35}
\vskip 1cm
\centerline{\Large \bf Mass, Entropy and Holography in}
\centerline{\Large \bf Asymptotically de Sitter Spaces}
\vskip 0.5 cm

\renewcommand{\thefootnote}{\fnsymbol{footnote}}
\centerline{{\bf Vijay
Balasubramanian,${}^{1}$\footnote{vijay@endive.hep.upenn.edu}
Jan de Boer,${}^{2}$\footnote{jdeboer@wins.uva.nl}
and
Djordje Minic${}^{3}$\footnote{dminic@vt.edu}
}}
\vskip .5cm
\centerline{${}^1$\it David Rittenhouse Laboratories, University of
Pennsylvania}
\centerline{\it Philadelphia, PA 19104, U.S.A.}
\vskip .5cm 
\centerline{${}^2$\it Instituut voor Theoretische Fysica,} 
\centerline{\it Valckenierstraat 65, 1018XE Amsterdam, The 
Netherlands} 
\vskip .5cm 
\centerline{${}^3$\it Institute for Particle Physics and Astrophysics}
\centerline{\it Department of Physics, Virginia Tech}
\centerline{\it Blacksburg, VA 24061, U.S.A.}
\vskip .5cm

\setcounter{footnote}{0}
\renewcommand{\thefootnote}{\arabic{footnote}}

\begin{abstract}
  We propose a novel prescription for computing the boundary stress
  tensor and charges of asymptotically de Sitter (dS) spacetimes from
  data at early or late time infinity.  If there is a holographic dual
  to dS spaces, defined analogously to the AdS/CFT correspondence, our
  methods compute the (Euclidean) stress tensor of the dual.  We
  compute the masses of Schwarzschild-de Sitter black holes in four
  and five dimensions, and the masses and angular momenta of Kerr-de
  Sitter spaces in three dimensions.  All these spaces are less
  massive than de Sitter, a fact which we use to qualitatively and
  quantitatively relate de Sitter entropy to the degeneracy of
  possible dual field theories. Our results in general dimension lead
  to a conjecture: {\it Any asymptotically de Sitter spacetime with
  mass greater than de Sitter has a cosmological singularity}.  
  Finally, if a dual to de Sitter exists, the trace of our stress
  tensor computes the RG equation of the dual field theory.
  Cosmological time evolution corresponds to RG evolution in the dual.
  The RG evolution of the c function is then related to changes in
  accessible degrees of freedom in an expanding universe.
\end{abstract}


\section{Introduction}
\label{intro}

There is no good local notion of energy in a gravitating spacetime. 
Nevertheless, there is a notion of mass, or total energy, which can be
computed from the effects of matter on spacetime geometry
\cite{wald}.  Heuristically, the deviation of the metric and
other fields near spatial infinity from their form in the vacuum
provides a measure of mass, angular momentum and other conserved
charges.  Equivalently, these charges can be computed from the
asymptotic symmetries of a spacetime; e.g., the eigenvalue of an
asymptotic timelike Killing vector will give a measure of mass.

There are two basic obstacles to applying this well-understood
philosophy to de Sitter space.  First, there is no spatial infinity --
the asymptotic regions of de Sitter are Euclidean surfaces at early
and late temporal infinity ($\CI^\pm$)~\cite{hawkell}.  Second, there
is no asymptotic Killing vector that is globally timelike.

In this article, we will evade these obstacles by computing the
quasilocal stress tensor of Brown and York~\cite{brownyork} on the
Euclidean surfaces at $\CI^{\pm}$, and by using this quantity to
define a novel notion of mass, and other charges appropriate to the
asymptotic symmetries.\footnote{Other interesting approaches to
defining mass in de Sitter space include~\cite{davjen}
and~\cite{confmass} which investigate possible positive mass
theorems.  The latter works define mass by using
timelike conformal Killing vectors of de Sitter space.}
Our methods are strongly reminiscent of the
computation of conserved charges in a conformal field theory.  If de
Sitter is dual to a Euclidean CFT~\cite{hull, hullkhuri, bhmds,
wittds, andyds} defined in a manner analogous to the AdS/CFT
correspondence~\cite{adsduality, adsduality2} we are computing the
energy and charges of states of the dual.

We use our stress tensor to display the asymptotic conformal
isometries of $\ds{3}$ and compute the central charge ($c$) of the
symmetry algebra.  In three dimensions, with a positive cosmological
constant, there are no black holes, but there is a spectrum of
spinning conical defects~\cite{park}.  We derive these solutions as
quotients of $\ds{3}$ and then compute their masses ($M$) and angular
momenta ($J$).  The conical defects are {\it less} massive than de
Sitter space.  Remarkably, naively entering $c$, $M$ and $J$ into the
Cardy formula of a hypothesized 2d CFT dual exactly reproduces the
entropy of the cosmological horizon of these spaces.  Related uses of
the Cardy formula to study $\ds{3}$ entropy appear in work by
Park~\cite{park} and, more recently, \cite{strombousso}.  Here we
point out several subtleties that face this reasoning, and some
potential resolutions.\footnote{Other approaches to de Sitter entropy
have been suggested in~\cite{carlip, banados, juanandy1, ent2}.}

We proceed to compute the masses of the Schwarzschild-de Sitter black
holes in 4 and 5 dimensions.  (See~\cite{klemm} for related
discussions.)  We find that de Sitter space is more massive than the
black hole spacetimes.  In fact, this is a qualitative feature
necessary for de Sitter entropy to have an interpretation in terms of
the degeneracy of a dual field theory defined in the AdS/CFT mode.  As
Bousso has emphasized, the entropy of de Sitter space is an upper
bound on the entropy of any asymptotically de Sitter
spacetime~\cite{boussobound}.  Since field theories generically have
entropies that increase with energy, a dual description of dS entropy
should map larger Schwarzschild-dS black holes into states of lower
energy, precisely as we find here.  Indeed, the largest black hole in
de Sitter space (the Nariai black hole~\cite{nariai}) has the least
mass.

Following Bousso~\cite{boussobound} we expect that asymptotically de
Sitter spaces with entropy greater than de Sitter do not exist. 
Therefore, since our mass formula increases monotonically with
entropy, we arrive at a conjecture:\footnote{As we will see below, we
could have chosen to define the classical stress tensor of de Sitter
space with the opposite sign.  With this definition, we would find
that de Sitter space has a lower mass than the black hole spacetimes
and our conjecture becomes a positive mass conjecture for non-singular
asympototically de Sitter spacetimes -- i.e., that de Sitter space
itself has the lowest mass among such spaces.}
\begin{center}
\begin{minipage}[c]{0.85\textwidth}
{\it
Any
asymptotically de Sitter space whose mass exceeds that of de Sitter
contains a cosmological singularity.
}
\end{minipage}
\end{center}

Given a holographic dual to asymptotically de Sitter space, the trace
of our stress tensor computes the dual RG equation.  We therefore
arrive at a remarkable picture: RG evolution of the dual is time
evolution in an expanding universe\footnote{As this paper was being
typed, the same point was made by Strominger \cite{andyrg}.}.  The
evolution of the central charge in the dual is directly related to the
changing number of accessible degrees of freedom in spacetime.  The
field theory c-theorem is related to properties of the Raychaudhuri
equation in gravity, in analogy with the properties of holographic RG
flows in the AdS/CFT correspondence \cite{fgpw, holorg, dvv, holorg1}.

We conclude the paper with a discussion of prospects for finding a
holographic dual to de Sitter space.  The material in this article has
been presented at a variety of conferences and workshops (see
Acknowledgments).  While the text was being prepared, we received a
number of preprints addressing various related 
aspects of de Sitter physics~\cite{otherdscft}. 

\section{Boundary stress tensor, mass and conserved charges}
\label{masssec}  

To date there is no completely adequate proposal for measuring the
mass of asymptotically de Sitter spaces.  The classic work of Abbott
and Deser remains the basic technique~\cite{abbottdeser}, but is a
perturbative approach measuring the energy of fluctuations. In AdS
space, the Brown-York boundary stress tensor~\cite{brownyork},
augmented by counterterms inspired by the duality with a
CFT~\cite{hensken,stressvp,emparan}, led to a particularly
convenient method of computing the mass of asymptotically AdS
spacetime.  Here we argue that a similar approach applies to de
Sitter space.

In $d+1$ dimensions, spaces with a positive cosmological constant 
solve  equations of motions derived from the action
\begin{equation}
    I_{B} = - {1 \over 16\pi G} \int_{\CM} d^{d+1} x \, \sqrt{-g} \, 
\left(
    R + 2\Lambda\right) + {1 \over 8\pi G} 
    \int^{\partial\CM^{+}}_{\partial\CM^{-}} d^d x \, \sqrt{h}  \, 
    K 
\labell{bareaction}
\end{equation}
Here $\CM$ is the bulk manifold, $\partial\CM^{\pm}$ are spatial
boundaries at early and late times, $g_{ij}$ is metric in the bulk of
spacetime, and $h_{\mu\nu}$ and $K$ are the induced metric and the
trace of the extrinsic curvature of the boundaries.  In de Sitter
space the spacetime boundaries $\CI^{\pm}$ are Euclidean surfaces at
early and late time infinity.   The notation
$\int^{\partial\CM^{+}}_{\partial\CM^{-}} d^2 x$ indicates an integral
over the late time boundary minus an integral over the early time
boundary which are both Euclidean surfaces.\footnote{We follow the
conventions of Brown and York~\cite{brownyork} for the boundary terms
and those of~\cite{stressvp} for the bulk term.} The extrinsic
curvature boundary term is necessary to allow a well-defined
Euler-Lagrange variation.  

It will be convenient to define a length scale
\begin{equation}
    l = \sqrt{{d(d-1) \over2\Lambda}} \, .
\end{equation}
In terms of $l$, the vacuum de Sitter solution to equations of motion 
derived from (\ref{bareaction}) is
\begin{equation}
    ds^{2} = -dt^{2} + l^{2} \cosh^{2}(t/l) \, (d\Omega^{2}_{d})
\labell{globmet}
\end{equation}
where equal time sections are $d$-spheres.  The same space admits a
coordinate system where equal time surfaces are flat:
\begin{equation}
ds^2 = -d\tau^2 + e^{2 \tau \over l} d\vec{x}^{2}
 =
{ l^{2} \over \eta^2} [ - d\eta^2 + d\vec{x}^{2}
] \, ,
\labell{infmet}
\end{equation}
with $\tau \in [-\infty, +\infty]$ while $\eta \in (0, \infty]$.  This
patch only covers half of de Sitter space, extending from a ``big
bang'' at a past horizon to the Euclidean surface at future infinity. 
By replacing $\tau$ by $-\tau$ a patch which covers the other half of
de Sitter (from past infinity to a future horizon) can be obtained. 
We will refer to these two patches as the ``big bang'' and the ``big
crunch'' patches.   Finally, an inertial observer in de Sitter space 
sees a static spacetime with a cosmological horizon:
\begin{equation}
         ds^{2} = - (1 - {r^{2} \over l^{2}}) \, dt^{2}
              + (1 - {r^{2} \over l^{2}})^{-1} \, dr^{2}
              + r^{2}\,d\Omega_{d-1}^{2} \, .
\labell{staticds}     
\end{equation}
The relations between these coordinate patches and Penrose diagrams 
are presented in Hawking and Ellis~\cite{hawkell} and references 
therein.   

We can formally ``Wick rotate'' from a positive to a negative
cosmological constant by the analytic continuation $l \rightarrow i
\,l$.  This formal transformation (sometimes accompanied by
additional Wick rotations of some of the coordinates) takes patches
of de Sitter into patches anti-de Sitter space.  For example, the
static path (\ref{staticds}) rotates to global AdS. Likewise, by
redefining $ e^{\tau/l} = r/l$ and carrying out some formal analytic
continuations gives the Poincar\'e patch of AdS. Tracking these
continuations through the classic computations of properties of AdS
spaces gives a powerful method of inferring some aspects of de Sitter
physics.

\paragraph{A finite action: } In general the action (\ref{bareaction})
is divergent when evaluated on a solution to the equations of motion,
because of the large volume at early and late times.  For example,
the late time region of the inflationary big bang patch makes a
divergent contribution to the action.  Specifically, if we cut time
off at a large time $t$, the leading term in the 3 dimensional 
action ($d+1 = 3$) is
\begin{equation}
    I = {1 \over 8\pi G} \int d^{2}x \, e^{2t/l} \, \left({-1 \over 
    l}\right) + {\rm finite}
    \labell{div1}
\end{equation}
which diverges as $t \rightarrow \infty$.  (The same applies to the
early time region of the inflationary big crunch patch.)  Mottola and
Mazur~\cite{mott} have shown that the results of Fefferman and
Graham~\cite{fefgrah} and Henningson and Skenderis~\cite{hensken} on
the asymptotics of spaces with a negative cosmological constant can be
extended to spaces with a positive cosmological constant. Combining
their results with the reasoning in~\cite{hensken, stressvp} shows
that the divergences of the action (\ref{bareaction}) can be canceled
by adding local boundary counterterms that do not affect the equations
of motion.

For example, in three dimensions the improved action
\begin{equation}
    I = I_{B} + {1 \over 8\pi G} \, \int_{\partial{\cal M}^{+}} 
d^{2}x 
    \sqrt{h} \,  {1 \over l}
    +  {1 \over 8\pi G} \, \int_{\partial{\cal M}^{-}}  d^{2}x 
    \sqrt{h}\, {1 \over l}
\labell{action}
\end{equation}
has the same solutions as (\ref{bareaction}) but is finite for
asymptotically de Sitter spaces.  Indeed, the counterterms in
(\ref{action}) clearly cancel the divergent terms of the bare action
in inflationary coordinates.  In analogy with AdS, if we place
boundary conditions requiring asymptotically dS spaces to approach the
de Sitter background sufficiently quickly at early and late times, the
divergences of the classical action will cancel.

In all dimensions the action has a class of divergences that are
powers of the conformal time coordinate $\eta$ appearing in
(\ref{infmet}).  In 3, 4 and 5 dimensions these are cancelled by the
counterterms
\begin{eqnarray}
   I_{{\rm ct}} &=& 
   {1 \over 8\pi G} \, \int_{\partial{\cal M}^{+}} d^{2}x \sqrt{h} \, 
   L_{{\rm ct}}
    +  {1 \over 8\pi G} \, \int_{\partial{\cal M}^{-}}  d^{2}x 
    \sqrt{h} \, L_{{\rm ct}} \labell{counteract} \\
    L_{{\rm ct}} &=&  {(d - 1) \over l} - {l^{2} \over  2(d-2)} R 
      \labell{counterlag}
\end{eqnarray}
The second counterterm only applies when $d+1 > 3$.  Here $R$ is the
intrinsic curvature of the boundary surface, and calculations are
performed by cutting off de Sitter space at a finite time, and then
letting the surface approach infinity.  In odd dimensions there is one
additional divergence that is logarithmic in the conformal time
$\eta$. This divergence cannot be cancelled without including an
explicit cutoff dependence in the counterterm action, thereby leading
to a conformal anomaly.  In the situations we study the anomaly and
the associated logarithmic divergence vanish and so we will neglect it
here.  See~\cite{hensken,haro} for a detailed discussion of this issue
in the framework of the AdS/CFT correspondence.  It would be
interesting to repeat the analysis of~\cite{pfr} to derive the
counterterms (\ref{counterlag}) for general dimensions from the
Gauss-Codazzi equations for spaces with a positive cosmological
constant.  Likewise, the analysis of logarithmic divergences
in~\cite{haro} should be extended to de Sitter.

\paragraph{Boundary stress tensor: } In AdS space, the Brown-York
prescription~\cite{brownyork,stressvp}
was used to compute a quasilocal boundary stress tensor that measures
the response of the spacetime to changes of the boundary metric.  We
can carry out an analogous procedure in de Sitter space to compute a
Euclidean boundary stress tensor on the spacetime boundary.
First, write the spacetime metric in ADM form as
\begin{equation}
    ds^{2} = g_{ij} \, dx^{i} \, dx^{j} =
    -N_{t}^{2} \, dt^{2} + h_{\mu\nu} \,
    (dx^{\mu} + V^{\mu} \, dt) \, (dx^{\nu} + V^{\nu} \, dt)
\end{equation}
Then $h_{\mu\nu}$ is the metric induced on surfaces of fixed time, 
and 
choosing $u^{\mu}$ to be the future pointing unit normal to these 
surfaces, 
we can compute the extrinsic curvature
\begin{equation}
    K_{\mu\nu} = - h_{\mu}^{i} \, \nabla_{i} 
    u_{\nu}
\end{equation}
and its trace $K$.  (Here the index on $h_{\mu\nu}$ is raised by the 
full 
metric $g_{ij}$.)  The Euclidean quasilocal stress tensor of 
de Sitter space is given by the response of the action, evaluated on 
the 
space of classical solutions, to variations of the 
boundary metric.   We can evaluate these variations either on an
early or late time boundary, getting the stress tensors
\begin{eqnarray}
T^{+\mu \nu} &=& {2 \over \sqrt{h}}
{ \delta I \over \delta h_{\mu \nu}} = \ \
- {1 \over 8\pi G} 
\left[ K^{\mu\nu} - K \, h^{\mu\nu} - {(d-1) \over l} \, 
h^{\mu\nu} - {l \over (d - 2)} \,G^{\mu\nu}
\right] \, , \labell{stress1} \\
T^{-\mu \nu} &=& {2 \over \sqrt{h}}
{ \delta I \over \delta h_{\mu \nu}} = \ \ 
- {1 \over 8\pi G} 
\left[ - K^{\mu\nu} + K \, h^{\mu\nu} - {(d-1) \over l} \, 
h^{\mu\nu} - {l \over (d - 2)} \,G^{\mu\nu}
\right] \, ,
\labell{stressminus}
\end{eqnarray} 
where the term proportional to $G^{\mu\nu}$, the Einstein tensor of
the boundary geometry, only appears when $d+1 > 3$.  The last two
terms in (\ref{stress1}) and (\ref{stressminus}) come from variation
of the counterterms in (\ref{counteract}).  To obtain the boundary
stress tensor we evaluate (\ref{stress1}) at fixed time and send the
time to infinity so that the surface approaches the spacetime
boundary.  The two equations (\ref{stress1}) and (\ref{stressminus})
appear to give different stress tensors because of the relative signs
of terms.  In fact, in empty de Sitter space they are evaluating
identical quantities -- the difference in signs occurs because the
extrinsic curvature $K$ is evaluated with respect to a future pointing
timelike normal leading to some sign changes between early and late
time surfaces.  For this reason, we will simply drop the $\pm$ and use
$T_{\mu\nu} \equiv T^{+\mu\nu}$ in the examples we study.  Note also
that there are some sign differences between (\ref{stress1}) and the
quasilocal stress tensor in AdS space~\cite{stressvp}.  These arise,
following Brown and York~\cite{brownyork}, from some differences
between the treatment of timelike and spacelike boundaries.

Since we are working on the Euclidean surface at $\CI^{\pm}$, we could
equally well have chosen to define the stress tensors in
(\ref{stress1}) and (\ref{stressminus}) with the opposite sign as
$(-2/\sqrt{h}) \, \delta I/\delta h_{\mu\nu}$.  This alternative sign
does not change any essential physics -- we will point out the
slightly different interpretations that follow in various places.

It is worth working out some examples of the stress tensor which we
will have use of later.  An equal time surface of the inflating patch
(\ref{infmet}) in 3 dimensions is the infinite Euclidean plane. 
Evaluating the stress tensor (\ref{stress1}) on this surface gives
\begin{equation}
    T^{\mu\nu}
     = - {1 \over 8\pi G l} e^{-2t/l} + {1 \over 8\pi G l} 
e^{-2t/l}
       =  0
\labell{stress2}
\end{equation}
Here the bare stress tensor canceled exactly against the counterterm.
By contrast, in global coordinates, the boundary stress tensor of
$\ds{3}$ is
\begin{equation}
    T^{\mu\nu} =  {e^{-t/l} \over 8\pi \, G \, l^{3} \, 
\cosh^{3}(t/l)}
         \pmatrix{1 & 0 \cr 0 & 1/\sin^{2}\theta} ~~~~~;~~~~~
         T = {1 \over 4\pi G l} \, {e^{-t/l} \over \cosh(t/l)} \, ,
         \labell{globstress}
\end{equation}
where $T$ is the trace of the stress tensor.  Notice that the stress 
tensor vanishes exponentially for $ t \to \infty$.

\paragraph{Mass and other conserved charges:} 
In a theory of gravity, mass is a measure of how much a metric
deviates near infinity from its natural vacuum behavior; i.e., mass
measures the warping of space.  The boundary stress tensor
(\ref{stress1}) computes the response of the spacetime action to such
a warping and thereby encodes a notion of mass.  Inspired by the
analogous reasoning in AdS space~\cite{brownyork, stressvp}, we 
propose a notion of mass for an asymptotically de Sitter space. 
We can always write the metric $h_{\mu\nu}$ on equal time surfaces in 
the
form:
\begin{equation}
    h_{\mu\nu} \, dx^{\mu} \, dx^{\nu } =
       N_{\rho}^{2} \, d\rho^{2} + 
       \sigma_{ab}\, (d\phi^a + N_\Sigma^a \, d\rho) \, 
               (d\phi^b + N_\Sigma^b \, d\rho)
       \labell{boundmet}
\end{equation}
where the $\phi^{a}$ are angular variables parametrizing closed
surfaces around an origin.  Let $\xi^{\mu}$ be a Killing vector
generating an isometry of the boundary geometry. 
Following~\cite{brownyork,stressvp} we can define the conserved charge
associated to $\xi^{\mu}$ as follows.  Let $n^{\mu}$ be the unit
normal on a surface of fixed $\rho$ and define the charge
\begin{equation}
   Q =  \oint_{\Sigma}  d^{d-1}\phi \,\sqrt{\sigma } \,   
   n^{\mu}\xi^{\mu} \,T_{\mu\nu}
   \labell{chargedef}
\end{equation}
In our computation $\rho$ will always be the coordinate associated
with the asymptotic Killing vector that is timelike inside the static
patch, but spacelike at $\CI^\pm$.
It would be interesting to compare this notion of a conserved charge 
in de Sitter space to that defined by Abbott and 
Deser~\cite{abbottdeser}.

An important difficulty facing the definition of mass in de Sitter
space is the absence of a globally timelike Killing vector.  However, as
is evident from (\ref{staticds}), there is a Killing vector that is
timelike inside the static patch, while it is {\it spacelike} outside
the cosmological horizon and therefore on $\CI^{\pm}$.  Any space that
is asymptotically de Sitter will have such an asymptotic symmetry
generator.  We can adapt the coordinates (\ref{boundmet}) so that
``radial'' normal $n^{\mu}$ is proportional to the relevant
(spacelike) boundary Killing vector $\xi^{\mu}$.  Then, we propose
that an interesting and useful notion of the mass $M$ of an
asymptotically de Sitter space is:
\begin{equation}
    M = 
    \oint_{\Sigma}  d^{d-1}\phi \,\sqrt{ \sigma } \, N_{\rho} \, 
\epsilon 
    ~~~~~;~~~~~ \epsilon \equiv
    n^{\mu}n^{\nu} \, 
    T_{\mu\nu} \, ,
    \labell{massdef}
\end{equation}
where we normalized the Killing vector in (\ref{chargedef}) as
$\xi^{\mu} = N_{\rho} n^{\mu}$.  Likewise, we can define momenta
\begin{equation}
    P_{a} = \oint_{\Sigma} \, d^{d-1}x \, \sqrt{\sigma} \, j_{a}
     ~~~~~;~~~~~ j_a = \sigma_{ab} \, n_{\mu} \, T^{a\mu} \, .
     \labell{angmom}
\end{equation}
We compute this formula on a surface of fixed time and then send time
to infinity so that it approaches the spacetime boundaries at
$\CI^{\pm}$.

Below we will investigate the meaning of the de Sitter stress tensor 
and conserved charges in various dimensions.

\section{Three dimensional cosmological spacetimes}

\subsection{More on Classical Solutions}

We seek an interesting class of solutions to 2+1 dimensional gravity
with a positive cosmological constant on which to test our stress
tensor and definition of mass.  In 3 dimensions black holes only exist
when there is a negative cosmological constant, but when $\Lambda > 0$
we can find a class of spinning conical defects which we will refer to
as the Kerr-de Sitter spaces following Park~\cite{park}.  These spaces
have been discussed before (see, e.g.,~\cite{deserjackiw, park,
 banados}), but we will derive them below as quotients of $\ds{3}$.
For convenience we will set the de Sitter scale to 1 ($l=1$) and will
restore it later by dimensional analysis.

Three dimensional de Sitter space is the quotient $SL(2,C)/SL(2,R)$,
and any solution of the field equations looks locally the same.  Thus
the general solution will locally look like $SL(2,C)/SL(2,R)$, but 
can be
subject to various global identifications.  In particular, we can
consider solutions of the form $\Gamma \backslash SL(2,C)/SL(2,R)$ 
for some
discrete subgroup $\Gamma$ of $SL(2,C)$.  If the discrete
subgroup is generated by a single element of $SL(2,C)$, there are two
possibilities.  Up to conjugation, the generator can be of the form
\begin{equation}
\left( \begin{array}{cc} q & 0 \\ 0 & q^{-1} \end{array} \right)
~~~~~~{\rm or}~~~~~~
\left( \begin{array}{cc} 1 & a \\ 0 & 1
\end{array}
\right) 
\end{equation}
The second case appears to correspond to solutions that look like the
inflationary patch (\ref{infmet}), but with the complex plane replaced
by a cylinder.  We will restrict our attention to solutions that
correspond to identifications of the first type.  

An element of $dS_3=SL(2,C)/SL(2,R)$ can be
written as
\begin{equation}
M= \left( \begin{array}{cc} u & i\alpha \\ i\beta & \bar{u} 
\end{array} \right),
\qquad u\, \bar{u}+\alpha \, \beta=1, \qquad \alpha,\beta \in R  \, ,
\label{matparam}
\end{equation}
in terms of which the metric is simply
\begin{equation}
ds^2 = du \, d\bar{u} + d\alpha \, d\beta \, .
\label{metricparam}
\end{equation}
$SL(2,C)$ acts on $M$ as
\begin{equation}
M \rightarrow \left(
\begin{array}{cc} a & b \\ c & d \end{array} \right) M \left(
\begin{array}{cc} \bar{d} & -\bar{b} \\ -\bar{c} & \bar{a} \end{array}
\right) 
\label{sl2param}
\end{equation}

Setting $q=\exp(\pi( r_- + i r_+))$, the identifications $\Gamma$
imposes on $dS_3$ are, in terms of the two by two matrix
parametrization, given by
\begin{equation}
\left( \begin{array}{cc} 
u & i\alpha \\ i\beta & \bar{u} \end{array} \right)  \sim
\left( \begin{array}{cc} 
e^{2 \pi i r_+} u & i e^{2 \pi r_-}
\alpha \\ i\beta e^{-2 \pi r_-} & \bar{u}
e^{-2 \pi i r_+}  \end{array} \right)  \, .
\end{equation}
To achieve such an identification, we parametrize the two by two
matrix as
\begin{equation}
\label{ident}
\left( \begin{array}{cc} 
u & i\alpha \\ i\beta & \bar{u} \end{array} \right) =
\left( \begin{array}{cc}
e^{i r_+ \phi} \sqrt{1-\alpha\beta} & i e^{r_- \phi}
\alpha \\ i\beta e^{- r_- \phi} & 
e^{- i r_+ \phi} \sqrt{1-\alpha\beta} \end{array} \right) 
\end{equation}
and take $\phi$ to be a periodic variable with period $2\pi$.  The
coordinates $\alpha,\beta$ have to satisfy $\alpha\beta \leq 1$.

The metric of (\ref{ident}) is of the form
\begin{equation}
ds^2 = (r_+^2 - \alpha\beta(r_+^2 + r_-^2))d\phi^2 + \ldots
\end{equation}
from which we see that in the region $\alpha\beta>r_+^2 (r_+^2 +
r_-^2)^{-1}$ the circle parametrized by $\phi$ becomes timelike. 
Therefore, we should remove this unphysical region.

The region $0\leq \alpha\beta \leq r_+^2 (r_+^2 +
r_-^2)^{-1}$ with $\alpha,\beta>0$ can be parametrized
as
\begin{equation}
\left( \begin{array}{cc} 
u & i\alpha \\ i\beta & \bar{u} \end{array} \right) =
\frac{1}{\sqrt{r_+^2 + r_-^2}}
\left( \begin{array}{cc}
e^{i (r_+ \phi+r_- t)} \sqrt{r^2 + r_-^2} &  i e^{r_- \phi-r_+ t}
\sqrt{r_+^2 - r^2} \\
i e^{- r_- \phi+ r_+ t} \sqrt{r_+^2 - r^2} & 
e^{- i ( r_+ \phi + r_- t)} \sqrt{r^2 + r_-^2} \end{array} \right) .
\end{equation}
The Kerr-dS metric derived from this reads
\begin{equation}
\label{metric2}
ds^2 = 
-\frac{(r^2 + r_-^2)(r_+^2 - r^2)}{r^2} dt^2 +
\frac{r^2}{(r^2 + r_-^2)(r_+^2 - r^2)}dr^2 +
r^2 (d\phi +\frac{r_+ r_-}{r^2} dt)^2 
\end{equation}
which looks very similar to that of the BTZ black hole \cite{btz} of 
3d gravity
with $\Lambda < 0$.  There is a horizon at $r=r_+$ of circumference
$2\pi r_+$.  The metric can be continued to $r^2 > r_+^2$, where the
roles of $t$ and $r$ are interchanged, so that the metric looks like
\begin{equation}
\label{metric4} ds^2 = -\frac{r^2}{(r^2 + r_-^2)(r^2-r_+^2)}dr^2 +
r^2 (d\phi +\frac{r_+ r_-}{r^2} dt)^2 + \frac{(r^2 +
r_-^2)(r^2-r_+^2)}{r^2} dt^2 \, .  
\end{equation}
At large $|r|$ this resembles the region close to ${\cal I}^\pm$ in
the Penrose diagram of $dS_3$.  Altogether, the Penrose diagram of the
Kerr de Sitter space is similar to the Penrose diagram of de Sitter
space, with static patches of the form (\ref{metric2}), and two
regions near past and future infinity.  Near ${\cal I}^{-}$ the metric
(\ref{metric2}) becomes \be ds^2 \sim -\frac{dr^2}{r^2} + r^2 (d\phi^2
+ dt^2) \ee and the spacelike slices are cylinders.  In contrast to
the case of $dS_3$, the points at $t\rightarrow \pm \infty$ of the
cylinder are not part of space-time.  The topology of Kerr de Sitter
space is $R^2 \times S^1$, whereas the topology of global de Sitter
space is $R \times S^2$.

The metric (\ref{metric2}) with $r_-=0$ 
\begin{equation}
    ds^{2} = -  (r_{+}^{2} - r^{2} ) \, dt^{2} + {dr^{2} \over 
(r_{+}^{2} - 
    r^{2})} + r^{2} \, d\phi^{2}
    \labell{conicdef}
\end{equation}
is a conical defect, with deficit angle $2\pi(1-r_+)$, describing a
world with a positive cosmological constant and a pointlike massive
observer.  When $r_+=1$ we reproduce global $dS_3$.  As in $\ads{3}$,
it is interesting task whether the spacetimes with $r_+>1$ (in effect,
``conical excesses'') make sense.  In $\ads{3}$, the AdS/CFT
correspondence tells us such spaces should be unphysical, so we might
expect that the situation in $\ds{3}$ is similar.

Another interesting limit is $r_+ \rightarrow 0$, where the deficit
angle becomes $2\pi$.  In this limit, the parametrization we used so
far is inadequate.  A convenient alternative is 
\begin{equation}
    \label{metric3}
\left( \begin{array}{cc} u & i\alpha \\ i\beta & \bar{u} \end{array}
\right) = \left( \begin{array}{cc} e^{i \psi} \sqrt{1+t^2} & i e^{r_-
\phi} t \\ i e^{- r_- \phi} t & e^{- i \psi} \sqrt{1+t^2} \end{array}
\right)  \, .
\end{equation}
Here, both $\psi$ and $\phi$ are periodic variables.  The
metric derived from (\ref{metric3}) reads \be ds^2 = -\frac{1}{1+t^2}
dt^2 + (1+t^2) d\phi^2 + r_-^2 d\psi^2 .  \ee The geometry represents
a torus that is contracted to a circle as $t\rightarrow 0$.  The
metric for $-\infty<t\leq 0$ is a kind of big crunch solution, the
metric for $0\leq t < \infty$ is a big bang type solution.  The
topology in each case is that of a solid two-torus, exactly the same
as that of the BTZ black hole.

More general big bang/big crunch solutions exist, with 
metric of a form identical to (\ref{metric4}),
\begin{equation} 
\label{metric5}
ds^2 = 
-\frac{t^2}{(t^2 + r_-^2)(t^2-r_+^2)}dt^2 +
t^2 (d\phi +\frac{r_+ r_-}{t^2} dr)^2 +
\frac{(t^2 + r_-^2)(t^2-r_+^2)}{t^2} dr^2 .
\end{equation}
(Here $r$ is periodic, and so the solution solution cannot be
continued through a horizon unlike (\ref{metric4}).)  It would be
interesting to see whether there exists a change of coordinates which
takes this metric to a form similar to the static patch of de Sitter
space.  An important difference is that we now require that both $r$
and $\phi$ are periodic variables.  Therefore, the metric cannot be
extended to $-r_+<t<r_+$.  For $t \leq -r_+$ the metric
(\ref{metric5}) is again some kind of big crunch solution, with a
torus contracting to a circle, whereas for $t \geq r_+$ it is a big
bang like solution.

The metric for the Kerr-de Sitter solution can be conveniently
rewritten as
\begin{equation}
ds^2 = - (8Gm - \frac{r^2}{l^2} + \frac{{(8GJ)}^2}{4r^2}) dt^2
+ (8Gm - \frac{r^2}{l^2} + \frac{{(8GJ)}^2}{4r^2})^{-1} dr^2
+ r^2 ( - \frac{8GJ}{2r^2} dt + d \phi )^2 \, .
\labell{kerrdsmet}
\end{equation}

\subsection{Mass and angular momentum of Kerr-dS spacetimes }

We begin by studying the Kerr-dS metric with $J=0$.  These are the
conical defect spacetimes appearing in (\ref{conicdef}), approaching
empty de Sitter when $8Gm = 1$.  For these spaces, there is a basic
subtlety in evaluating the boundary stress tensor of $\ds{3}$ that
enters the candidate mass formula (\ref{massdef}).  In the region $r <
l\sqrt{8Gm}$, equal $t$ surfaces of (\ref{kerrdsmet}) approach the
cosmological horizon (rather than $\CI^{\pm}$ at early and late times).
By contrast, when $r>l\sqrt{8Gm}$, $r$  becomes timelike while $t$ is a
spatial direction, and large $r$ surfaces approach $\CI^{\pm}$.  Since
we propose to define the stress tensor and mass at $\CI^{\pm}$, we
will evaluate these quantities on surfaces of large fixed $r$ which
therefore lie {\it outside} the cosmological region accessible to the
observer creating the conical defect in (\ref{conicdef}).

An equal time surface of the conical defect spacetimes outside the
cosmological horizon ($r > l\sqrt{8Gm}$) has a metric
\begin{equation}
    ds^{2} = ({r^2 \over l^{2}} - 8Gm) \, dt^{2} + r^{2} \, 
    d\theta^{2} \, .
\end{equation}
We can evaluate the stress tensor (\ref{stress1}) on this surface and
compute the mass (\ref{massdef}); each step is almost identical to the
analogous AdS computations~\cite{stressvp}.  As $r \rightarrow \infty$
(recall $r$ is timelike now) we find
\begin{equation}
    M = {1 \over 8\pi G} \oint d\theta \, {8Gm \over 2} = 
    {8Gm  \over 8 G} = m
\end{equation}
Setting $8Gm = 1$ we find that $\ds{3}$ is assigned a mass of 
$1/8G$.   

Surprisingly, we are finding that the conical defects have {\it lower}
mass than pure de Sitter space.  We might interpret this as follows. 
Even if the matter making up the defects itself has positive energy,
the binding energy to the gravitational background can decrease the
total energy.  In particular, a conical defect ``swallows up'' a part
of the spacetime and thereby appears to reduce the net amount of
energy stored in the cosmological constant. 

The computation of the Brown-York boundary stress tensor
for the general Kerr-de Sitter spacetime parallels the analogous
computation for the BTZ black hole~\cite{stressvp}.  Specifically, if
\begin{equation}
ds^2 = -\frac{l^2}{r^2} dr^2 + \frac{r^2}{l^2}(d\tau^2 +  dx^2)
+ \delta g_{MN} dx^M dx^N
\end{equation}
one finds that the mass and the angular momentum are given by the
following expressions
\begin{equation}
M = \frac{l}{8\pi G} \int dx \, \left[ \frac{r^4}{2l^5} \delta g_{rr}
+ \frac{1}{l} \delta g_{xx} - \frac{r}{2l} \partial_r \delta 
g_{xx}
\right]
\end{equation}
and
\begin{equation}
P_{x} =  - \frac{l}{8\pi G} \int dx \,  \left[ 
\frac{1}{l} \delta g_{\tau x} - \frac{r}{2l} \partial_r \delta 
g_{\tau x}
\right]
\end{equation}
where in the case of the Kerr-de Sitter solution
\begin{equation}
\delta g_{rr} = \frac{8Gml^4}{r^4}, \quad \delta g_{\tau \tau} = 8Gm,
\quad  \delta g_{\tau \phi} = -4GJ
\end{equation}
with $x = l \phi$ and $\phi \in [0, 2\pi]$.  Thus the final result is
\begin{equation}
    M = m ~~~~~;~~~~~ P_{\phi} = J \, .
\end{equation}
As a cross-check, note that we recover the mass $M=\frac{1}{8G}$ of de
Sitter if we set $J=0$ and $m=\frac{1}{8G}$.

Our computation of the mass and angular momentum of the Kerr-dS 
spaces is strongly reminiscent of the techniques of Euclidean 
conformal 
field theory.  The massive spinning defect intersects a point on the 
Euclidean surface at $\CI^{+}$  which is inside the cosmological 
horizon and excised from the exterior region.  We find the charges 
carried by the defect performing a contour integral around this 
excised point.     This computation mimics the usual CFT procedure of 
inserting an operator at the origin around which we integrate to 
compute charges.

\section{Conformal symmetries and entropy of $\ds{3}$}
\label{conformal}

We will now employ the boundary stress tensor developed in the
previous section to the study of the symmetry group of asymptotically
$\ds{3}$ spaces.  We will first study the $SL(2,C)$ isometry group and
then the group of asymptotic conformal symmetries.

\subsection{Isometries}
\label{isomet}
We will set the de Sitter scale $l$ to $1$, restoring it as needed by
dimensional analysis.  In (\ref{matparam},\ref{metricparam}) we
indicated how $\ds{3}$ can be represented as the group manifold
$SL(2,C)/SL(2,R)$. We can represent each of the three metrics for de
Sitter that we discussed earlier (inflating, global and static) in
terms of the matrix $M$.  Global coordinates for $\ds{3}$
(\ref{globmet}) correspond to parametrizing $M$ as
\begin{equation}
M= \left( \begin{array}{cc}
\cosh t \, \sin \theta \, e^{i\phi} & i ( \sinh t + \cosh t \,  \cos 
\theta) \\
i ( -\sinh t + \cosh t \,  \cos \theta) & \cosh t \,  \sin \theta \,  
e^{-i \phi}
\end{array} \right) \, ,
\end{equation}
with induced metric
$
ds^2 = -dt^2 + \cosh^2 t (d\theta^2 + \sin^2 \theta d\phi^2) 
$.
The big bang inflationary patch (\ref{infmet}) corresponds to 
$\alpha>0$. 
This is parametrized as
\begin{equation}
M= \left( \begin{array}{cc}
 e^t z & ie^t \\ i(e^{-t} - e^t z \bar{z}) & e^t \bar{
z} \end{array} \right) ~~~~~
\Longrightarrow ~~~~~ ds^2 = -dt^2 + e^{2t} dz d\bar{z} .
\end{equation}
Similarly, one can parametrize the big bang/big crunch and 
inflationary patches
corresponding to $\alpha<0$, $\beta>0$ and $\beta<0$.  A static patch
(\ref{staticds}) is covered by $\alpha>0,\beta>0$. 
The matrix $M$ is parametrized as
\begin{equation}
M=\left( \begin{array}{cc} e^{i\phi} r & i e^t \sqrt{1-r^2} \\
  i e^{-t} \sqrt{1-r^2}  & e^{-i\phi} r \end{array} \right) 
  ~~~\Longrightarrow~~~
  ds^2 = -(1-r^2) dt^2 +\frac{1}{1-r^2} dr^2 + r^2 d\phi^2  \,.
\end{equation}

\paragraph{$SL(2,C)$ Action: }
One convenient feature of the parametrization in terms of the matrix
$M$ is that it is easy to describe action of the $SL(2,C)$ isometry
group on the coordinates via (\ref{sl2param}).  From this we get the
relevant Virasoro action in the different patches.  Notice that the
actual real generators are of the form $\alpha L_i + \bar{\alpha}
\bar{L}_i$, where $\bar{L}_i$ is the complex conjugate of $L_i$.  For
the global patch we get
\begin{eqnarray}
    L_{-1} &
= & \frac{i}{2} e^{-i\phi} (\cot\theta + \frac{\tanh t}{\sin \theta} )
\partial_{\phi} - \frac{1}{2} e^{-i \phi} \sin \theta
\partial_{\theta} \nonumber \\
& & -\frac{1}{2} e^{-i \phi} (\cosh t + \cos \theta \sinh t) 
\partial_t 
\labell{globgenmone} \\
L_0 & = & \frac{i}{2} \partial_{\phi} + \frac{1}{2} \cos \theta 
\partial_{\theta} 
-\frac{1}{2} \sin \theta \tanh t \partial_t 
\labell{globgenzero} \\
L_{1} & = & -\frac{i}{2} e^{i\phi} (\cot\theta - \frac{\tanh t}{\sin 
\theta} ) \partial_{\phi} +
\frac{1}{2} e^{i \phi} \sin \theta \partial_{\theta} \nonumber \\
& & +\frac{1}{2} e^{i \phi} (-\cosh t + \cos \theta \sinh t) 
\partial_t 
\labell{globgenone}
\end{eqnarray}
In the inflationary patch we obtain
\begin{eqnarray}
L_{-1} & = & -\partial_z \labell{flatgenmone} \\
L_0 & = & \frac{1}{2} \partial_t - z \partial z \labell{flatgenzero} 
\\
L_1 & = & z \partial_t - z^2 \partial_z - e^{-2 t} \partial_{\bar{z}}
\labell{flatgenone}
\end{eqnarray}
and in the static patch
\bea
L_{-1} & = & -\frac{1}{2r} (1-r^2)^{1/2} e^{t-i\phi} (\partial_t -2 r 
\partial_r)
 + \frac{ir}{2} (1-r^2)^{-1/2} e^{t-i\phi} \partial_\phi 
 \labell{staticgenmone} \\
L_0 & = & \frac{1}{2} \partial_t + \frac{i}{2} \partial_{\phi} 
\labell{staticgenzero} \\
L_{1} & = & \frac{1}{2r} (1-r^2)^{1/2} e^{-t+i\phi} (\partial_t +2 r 
\partial_r)
 - \frac{ir}{2} (1-r^2)^{-1/2} e^{-t+i\phi} \partial_\phi \, .
 \labell{staticgenone}
\end{eqnarray}
In particular, we see that in the static patch, $L_0+\bar{L}_0 
=\partial_t$.

\subsection{Asymptotic conformal symmetry}
Brown and Henneaux~\cite{bh} specified boundary conditions for
asymptotically $\ads{3}$ spaces that admitted a well defined algebra
of diffeomorphisms.  These continue to boundary conditions at future
or past infinity of $\ds{3}$, defining an asymptotically de Sitter
geometry.  Working in the inflating coordinate system (\ref{infmet})
we obtain that an asymptotically $\ds{3}$ space has a metric that
satisfies \cite{andyds}
\begin{equation}
g_{+-} = -{ e^{2 \tau \over l} \over 2} + O(1), \quad g_{++}, 
g_{--} = O(1), 
\quad
g_{\tau \tau} = -1 + O(e^{-{2\tau \over l}}) 
\quad
g_{+r}, g_{-r} = O(e^{-{3\tau \over l}}) \, . \labell{bc}
\end{equation}
The diffeomorphisms that respect these conditions are given in terms 
of functions $\xi^+(z^+)$ and $\xi^{-}(z^-)$, and take the asymptotic 
form \cite{andyds}:
\begin{eqnarray}
z^+ &\rightarrow& z^+ - \xi^{+} -
{l^2 \over 2} \,  e^{- {2 \tau \over l}} \,  \partial_{-}^2 \xi^{-}
\labell{diff1} \\
z^- &\rightarrow& z^- - \xi^{-} -
{l^2 \over 2} \, e^{- {2 \tau \over l}} \, \partial_{+}^2 \xi^{+}
\labell{diff2}
\\
e^{\tau \over l} &\rightarrow& e^{\tau \over l} +
{e^{\tau \over l} \over 2} (\partial_{+}\xi^{+} +\partial_{-} 
\xi^{-}\, )
\labell{diff3}
\end{eqnarray}
To understand the meaning of these diffeomorphisms, consider the
metric $\gamma_{\mu\nu}$ induced on an equal time surface of large
$\tau$ in (\ref{infmet}).  To leading order in $\tau$, (\ref{diff1})
and (\ref{diff2}) produce a conformal transformation of this metric.
The resulting conformal factor in $\gamma_{\mu\nu}$ is undone by the
effective Weyl transformation induced by (\ref{diff3}), leaving the
leading asymptotic metric unchanged \cite{andyds}:
\begin{equation}
    ds^{2} \rightarrow - d\tau^{2} + e^{2\tau \over l} \, 
dz^{+}\,dz^{-} 
    + {l^{2} \over 2} \, (\partial_{+}^{3} \xi^{+}) (dz^{+})^{2}
    + {l^{2} \over 2} \, (\partial_{-}^{3} \xi^{-}) (dz^{-})^{2}
\labell{transfmet}
\end{equation}
We learn from this that the asymptotic symmetry group of $\ds{3}$,
subject to the boundary conditions (\ref{bc}), is the two dimensional
Euclidean conformal group, which contains the isometries discussed
above as a subgroup.  Indeed, the detailed analysis of Brown and
Henneaux~\cite{bh} can be formally analytically continued to arrive 
at a
Virasoro symmetry algebra, but we will not present the details here. 
Clearly, a similar analysis can be carried out to show that there is a
conformal group of asymptotic symmetries at past infinity also.

In summary, the asymptotic symmetry group of de Sitter space is the
conformal group.  Interestingly, time translation in the inflating
patch of de Sitter space is related to the dilatation generator of
the conformal group.  Explicitly, the isometry generator 
(\ref{flatgenzero}) in the static patch shows that a dilatation of  
the boundary can be undone by a time translation.  It is also 
illuminating to rewrite the isometry generators 
(\ref{flatgenmone})--(\ref{flatgenone}) using the conformal time 
introduced in (\ref{infmet}).  They become:
\begin{equation}
    L_{-1} =  -\partial_{z} ~~~~~;~~~~~ L_{0} =  -{\eta\over 2} \, 
\partial_{\eta}
       - z\partial_{z} ~~~~~;~~~~~
    L_{1} = -z\eta\,\partial_{\eta} - z^{2}\partial_{z} - \eta^{2} 
    \partial_{\bar{z}}
\labell{conftimegens}    
\end{equation}
It is then clear that as $\eta \rightarrow 0$, reaching the boundary
of de Sitter, the isometries reduce to standard conformal
transformations of the plane, while $L_{0}$ and the conjugate
$\bar{L}_{0}$ generate dilatations of the boundary.  As we will see
later, if there is a holographic dual to de Sitter space, this leads
to a correspondence between the renormalization group of the dual and
time translation in de Sitter.

We will compute the central charge of the asymptotic conformal
symmetry using the methods of~\cite{stressvp}.  We have shown in this
section that there is an asymptotic group of conformal symmetries.
However, as we will see, the action (\ref{action}) is not left
invariant by these conformal transformations; there is an anomaly
produced by the procedure of cutting off the space to regulate
divergences and then removing the cutoff. (This anomaly is intimately
connected to the logarithmic divergences discussed below
(\ref{counterlag}).  Such divergences and the resulting conformal
anomlies were discussed for asymptotically AdS spaces
in~\cite{hensken,haro}.) The stress tensor (\ref{stress1}) computes
the change in the action to variations of the boundary metric.  Hence
we can measure the central charge of the conformal symmetry group of
$\ds{3}$ by computing the anomalous transformation of this stress
tensor under the diffeomorphisms (\ref{diff1})--(\ref{diff3}), which
produce the transformed metric (\ref{transfmet}). We showed in
(\ref{stress2}) that the boundary stress tensor of the inflating patch
vanishes.   After the diffeomorphisms (\ref{diff1}--\ref{diff2}), the
boundary stress  tensor becomes
\begin{equation}
 T_{++} = -{l \over 16 \pi G} \partial_{+}^3 \xi{+}, \quad
T_{--} = -{l \over 16 \pi G} \partial_{-}^3 \xi{-} \, .
\end{equation}
Using the standard formulae for the anomalous transformation of the
stress tensor in a two dimensional conformal theory, we read off the
central charge
$
c={-3l \over 2G} \, ,
$
which has the same formal dependence on the cosmological length scale
as the central charge of $\ads{3}$.  The negative sign of $c$ is not
problematic here since we are dealing with a classical stress tensor. 
If, as metioned before, we had chosen to define a stress tensor with
the opposite overall sign as $(-2/\sqrt{h}) \delta I/\delta
h_{\mu\nu}$, the resulting object would of course have a positive
central charge under conformal transformations.  The same result for
the central charge is obtained by examining the trace of the stress
tensor in global coordinates as $t\rightarrow\infty$
(\ref{globstress}).  Remembering that $T = - (c/24\pi) R$, where $R$
is the scalar curvature of the spherical boundary, we again find $c =
-3l/2G$.

The analysis outlined above can be repeated in four and five
dimensions.  We can compute the Brown-York stress tensor on the
spacetime boundary, and the anomalous variation of the this tensor for
five dimensional de Sitter will yield a central charge.  However, the
absence of such an anomaly for four dimensional de Sitter prevents use
of this tool in that case.  All computations proceed in analogy to
those in~\cite{stressvp}.

\subsection{Brown-York, Cardy and Bekenstein-Hawking}

The Kerr-dS spacetimes ((\ref{kerrdsmet}) have
cosmological horizons giving rise to a Bekenstein-Hawking entropy 
\cite{bekhaw, gh1, gh2}
\begin{equation}
S_{BH}^{K-dS}= {\sqrt{2} \pi l \over 4 G} 
\sqrt{ (8Gm) +
\sqrt{(8Gm)^2 + \frac{(8GJ)^2 }{l^2} } }
\labell{kerrentropy}
\end{equation}
In particular when $J=0$, the horizon is at $r=l\sqrt{8Gm}$ and the
entropy is $S= {\pi l\sqrt{8Gm} \over 2G}$.  As illustrated in
previous sections (also see \cite{mott, wittds, andyds, klemm}) there
is an asymptotic conformal algebra in $\ds{3}$ in detailed analogy
with the $\ads{3}$ symmetry algebra uncovered by Brown and Henneaux
\cite{bh}.  Therefore, if there is a holographic dual to de Sitter
space, we might expect it to be a Euclidean conformal field
theory~\cite{wittds, andyds}.  Alternatively, the dual theory might be
in fact a Lorentzian conformal field theory with a Euclidean signature
Virasoro algebra.  In addition, Strominger has emphasized that the
results of Brown and Henneaux can be understood as saying that states
of any well defined quantum gravity on $\ads{3}$ transform in
representations of the conformal group, and that the same applies to
$\ds{3}$.

If there is a duality between de Sitter space and a CFT it is not yet
on a solid footing -- we neither know how to realize de Sitter space
in string theory, nor how the gravitational data could be related to a
dual.  However, it might be natural to suppose that in analogy with
AdS, the dS stress tensor is related to the stress tensor of a dual. 
Indeed, the actual definition of spacetime conserved charges in
(\ref{chargedef},\ref{massdef},\ref{angmom}) is in formal analogy to
the standard definition of conserved charges in a Euclidean CFT. In
effect our treatment excises a point on $\CI^{+}$ where the static
patch meets future or past infinity and the contour integral in
(\ref{chargedef}) is carried out around this point.  We might imagine
that from the perspective of a dual theory the operator responsible
for creating the spacetime state is placed at this excised point and
our formula computes the charges of the state after mapping the plane
to the cylinder.

However, in making such an identification there are several subtle
issues.  First of all, there are various possible sign flips that may
be relevant.  For example, if we want to follow the radial
quantization analogy given above, the Euclidean time coordinate
obtained by continuation out of the static patch flows in the wrong
direction at $\CI^{+}$.  (After mapping to the plane it flows towards
the radial origin where the operator on $\CI^{+}$ would be inserted
rather than towards radial infinity.)  For similar reasons, CFT stress
tensors on $\CI^{+}$ and $\CI^{-}$ that describe the same space might
be related to our dS stress tensor with opposite signs.  Issues of
this kind can be solved properly if an actual technical definition of
a dS/CFT correspondence is devised.  Here we content ourselves with
the following tempting numerological observation.

The considerations above, coupled with Strominger's well-known
observations regarding the entropy of the BTZ black
hole~\cite{andybtz}, suggest that the entropy of Kerr-dS spaces could
be explained by applying a Cardy-like formula to a CFT with \ left and
right energy levels measured by the eigenvalues of the $L_{0}$ and
$\bar{L}_{0}$ conformal generators in static coordinates, as defined
in (\ref{staticgenzero}).  From their definition, we see that these
$L_{0}$ and $\bar{L}_{0}$ eigenvalues are related to the mass
(\ref{massdef}) and angular momentum (\ref{angmom}) by
\begin{equation}
    L_{0} + \bar{L}_{0} = l \,m
    ~~~~~;~~~~~
    L_{0} - \bar{L}_{0} = i J \, .
\end{equation}
Formally, this gives
\begin{equation}
    L_0  = \frac{1}{2} (ml + iJ) ~~~~~;~~~~~
    \bar{L}_0 =\frac{1}{2} (ml - iJ) \, .
\label{Leigvals}
\end{equation}
(Also see~\cite{andyds} for details relating the Brown-York stress
tensor to the Virasoro generators.)  When $J=0$ naive application of
a Cardy-like formula for the asymptotic density of states of a unitary
CFT
\begin{equation}
S_C= 2\pi \sqrt{{|c|L_0 \over 6}} +2\pi \sqrt{{|c|\bar{L}_0
\over 6}} 
\labell{cardy}
\end{equation}
gives the entropy
\begin{equation}
S = {\pi l \sqrt{8Gm} \over 2G}
\end{equation}
 in exact agreement with the entropy of the conical defects
 (\ref{conicdef}) (also see~\cite{strombousso}).\footnote{We have
 placed absolute value signs around $c$ because in our definition of
 the spacetime stress tensor $c<0$.  As we mentioned above there are
 several subtle issues in relating the signs of the spacetime stress
 tensor to possible duals and one hopes that these issues could
 resolved given a technical definition of a full-fledged duality. 
 Regardless, the numerological observation made here remains
 interesting.}

Now observe that when that the spinning Kerr-dS spacetimes give
rise to complex eigenvalues for $L_{0}$ and $\bar{L}_{0}$.  A CFT
whose left and right energy levels these are therefore cannot be
unitary.  Nevertheless, we naively apply  a Cardy-like
formula (\ref{cardy}) and find that:
\begin{eqnarray}
    S &=& 
    2\pi \sqrt{{|c| (ml + i J) \over 12}} +2\pi \sqrt{{|c| (m l - i J) 
    \over 12}} \\  
    &=& {\sqrt{2} \pi l \over 4 G}
\sqrt{ (8Gm) +
\sqrt{(8Gm)^2 + \frac{(8GJ)^2 }{l^2} } }
\end{eqnarray}
The complex $L_{0}$ eigenvalues suggest that the Cardy formula cannot
be valid, since it does not generally apply to non-unitary theories,
but we have nevertheless exactly reproduced the Kerr-dS entropy
(\ref{kerrentropy}).

In fact, there is a further subtlety.  The complete Cardy formula for
the asymptotic density of states of a unitary 2d CFT is:
\begin{equation}
S_C= 2\pi \sqrt{{c\over 6} (L_{0} - c/24)} +2\pi \sqrt{{c \over 6} \,
(\bar{L}_0 - c/24)}    
\end{equation}
and we might expect large corrections when $L_{0} , \bar{L}_{0} \sim
c/24$.  For $\ds{3}$, $M = 1/8G$ and so $L_{0} = \bar{L}_{0} = 1/16 G
= |c|/24$ and so it is surprising that dropping the $|c|/24$ shift
still gives the ``right'' answer for the entropy.  The important point
is that the entropy of $\ds{3}$ (and of the conical defects) scales
linearly with the central charge.  We have in fact found that in
generic CFTs with a central charge $c$, when $L_{0} \sim c/24$, the
degeneracy of states behaves as
\begin{equation}
    d \sim e^{\lambda c}
\end{equation}
with a non-universal coefficient $\lambda$.   Also, if the CFT dual 
to $\ds{3}$ is related to the D1-D5 CFT which was dual to $\ads{3}$, 
one might be able to use U-duality to show that the formula 
(\ref{cardy}) applies even when $L_{0} \sim c/24$.

Actually, in the above discussion we have been somewhat imprecise
regarding the hermiticity conditions and reality conditions on $L_k$
and $\bar{L}_k$.  As one can see from
(\ref{staticgenmone})-(\ref{staticgenone}), the $L_k$ satisfy the
hermiticity condition $L_k^{\dagger}=-\bar{L}_k$.  On the other hand,
the eigenvalues given in (\ref{Leigvals}) are not compatible with this
hermiticity condition.  The reason is that conventionally, the energy
or mass is the eigenvalue associated to the operator $-i\partial_t$,
while the mass is computed from the Brown-York tensor using the
$\partial_t$ Killing vector.  In our case, the operator $-i\partial_t$
is $-i(L_0+\bar{L}_0)$, as is clear from
(\ref{staticgenmone})-(\ref{staticgenone}).  Therefore, we could have
identified $lm$ with $-i(L_0+\bar{L}_0)$, and $iJ$ with
$-i(L_0-\bar{L}_0)$, which gives $L_0=\frac{i}{2} (ml+iJ)$ and
$\bar{L}_0=\frac{i}{2}(ml-iJ)$.  At the same time, the modes of the
stress tensor as we defined it would be identified with $L_k$, but
rather with $iL_k$.  This implies that the central charge should have
been equal to $-3il/2G$.  Putting these values for $L_0,\bar{L}_0$ and
$c$ into the Cardy formula reproduces the Kerr-dS entropy; they are
also identical to the values obtained from a Chern-Simons theory
analysis by Park~\cite{park}.  The central lesson to be learned here
is that while relating a Lorentzian bulk to a Euclidean boundary there
are factors of $i$ and unusual reality conditions which will be
important to understand for the definition of a possible dS/CFT
correspondence.

It is interesting to note that that similar conclusions can be reached 
by naively continuing the AdS results in~\cite{andybtz} to dS by 
complexifying the scale: $l \rightarrow il$.  The central charge and 
$L_{0}$ eigenvalues become imaginary, but conspire to correctly give 
the entropies of the de Sitter conical defects in a naive application 
of the Cardy formula.

Finally, the expression for the Brown-York mass of a conical defect
allows us to compute the Hawking temperature from the first law
of thermodynamics 
\begin{equation}
dE  = T dS \, ,
\end{equation}
and the fact that $E=\frac{\gamma^2}{8G}$, while the entropy is $S=
\frac{\gamma \pi l}{2G}$.  (We have written $\gamma^{2} = 8Gm$.  By
considering variations over $\gamma$ we deduce that
\begin{equation}
T_H = \frac{dE}{dS} = \frac{\gamma}{2\pi l}
\end{equation}
as it should be \cite{nappi, gh1, gh2}. In the limit when $\gamma = 1$
we obtain the correct expression for the Hawking temperature of de 
Sitter space.

Banks has emphasized that the finite entropy of de Sitter leads us
expect that quantum gravity in a universe with a positive cosmological
constant has a finite number of degrees of freedom \cite{banks1, bf}. 
It is worth examining how this might be reconciled with the
proposition that a Euclidean conformal field theory may be dual to de
Sitter, since any local field theory has an infinite number of degrees
of freedom.  The picture emerged above is that de Sitter space should
be described by an ensemble of states with $L_{0} = {\bar{L}}_{0}=
|c|/24$.  The finite degeneracy of such states would account for de
Sitter entropy.  (This is reminiscent of \cite{stromvaf}.)  Likewise,
states of lower $L_{0}$ contribute to ensembles that describe the
classical conical defects.  It this picture it is natural that the
conical defects and Kerr-dS spaces have masses smaller than that of
$\ds{3}$.

But what about states with the real part of $L_0 > |c|/24$? These
correspond to a spacetimes with mass bigger than that of $\ds{3}$.
Consider such a universe, which is asymptotically $\ds{3}$ at early
times and in which the mass formula (\ref{massdef}) at $\CI^{-}$
measures a mass $M > 1/8G$. It is likely that such a space evolves to
a singularity and is not asymptotically de Sitter in the future.
Hence, most states, having ${\rm Re}(L_{0}) > |c|/24$, would not lead to de
Sitter-like evolution and therefore the specification of de Sitter
boundary conditions would restrict the dual theory to a finite number
of states. In turn this imply that the conical parameter $8 G m \le
1$.

\section{Higher dimensional de Sitter and a mass bound}

We can test the picture of de Sitter entropy that emerged in the 
previous section by examining what our mass formula says about black 
holes in higher dimensions.   When $d + 1 > 3$ the action 
(\ref{bareaction}) admits Schwarzschild-dS black hole solutions of 
the 
form:
\begin{eqnarray}
    ds^{2} &=& - V(r) \, dt^{2} + V(r)^{-1} \, dr^{2} + r^{2} \, 
    d\Omega_{d-2} \, , \labell{SdS} \\
    V(r) &=& 1 - {2Gm \over r^{d-2}} - {2\Lambda r^{2} \over d(d-1)}
             = 1 - {2Gm \over r^{d-2}} - {r^{2} \over l^{2}} \, .
\end{eqnarray}
This space has horizons at locations where $V(r) = 0$.  

When $m=0$ we recover static coordinates for empty de Sitter space
with a single cosmological horizon at $r=l$ and an entropy $S =
l^{d-2} V_{d-2}/4G$ where $V_{d-2}$ is the volume of the unit
$(d-2)-$sphere.  As $m$ increases, a black hole horizon appears,
increasing in size with $m$.  Simultaneously the cosmological horizon
shrinks in size, pulled inwards by the gravitational attraction of the
black hole.  As a result there is a largest black hole, the Nariai
solution, which occurs when
\begin{equation}
    m = m_{N} = \left({1 \over d G}\right) \, 
    \left[ {(d-1)(d-2) \over 2\Lambda} \right]^{(d-2)/2}.
\end{equation}
For this critical choice of mass, the black hole and cosmological
horizons coincide at a radius $r^{2} = (d-1)(d-2)/2\Lambda = l^{2}
(d-2)/d$, giving rise to a Bekenstein-Hawking entropy of $S = l^{d-2}
V_{d-2} (1 - 2/d)^{(d-2)/2} /4G$.  Spaces with $m > m_{N}$ have
unacceptable naked singularities.  The Nariai solution can also be
reparametrized as (e.g., see (\cite{ginsperry}):
\begin{eqnarray}
    ds^{2} &=& - W(r) \, dt^{2} + W(r)^{-1} \, dr^{2} +  {(d-1)(d-2) 
    \over 2\Lambda} \, 
    d\Omega_{d-2} \, ,  \\
    W(r) &=& 1 - {2\Lambda \, r^{2} \over (d-1)}
             = 1 -  {d \, r^{2} \over l^{2}} \, .
\end{eqnarray}

As we noted above, the Nariai solution has a smaller entropy than 
that 
of de Sitter space.  Indeed, the sum of the entropies of black hole 
horizons, cosmological horizons and matter fields cannot exceed the 
entropy of de Sitter space, given suitable energy conditions on the 
matter~\cite{boussobound}.   

We can compute the masses of the black holes (\ref{SdS}) using 
(\ref{massdef}).   In four dimensions we find that
\begin{equation}
    M_{4} = -m \, .
    \labell{4dmass}
\end{equation}
So $\ds{4}$ has a vanishing mass according to our formula while the 
Nariai black hole has $M_{4} = - (1/3G\sqrt{\Lambda})$.  In five 
dimensions the mass formula becomes
\begin{equation}
    M_{5} =  {3\pi l^{2} \over 32 G} - {3\pi m \over 4}\, .
\labell{5dmass}
\end{equation}
(Note that this result differs in sign from~\cite{klemm}.)  So
the mass of $\ds{5}$ is $2\pi l^{2}/32 G$, and, pleasantly, the mass
of the 5d Nariai black hole is 0.  This is parallel to the three
dimensional case where the biggest conical defect had a mass of zero.

If there is a CFT dual to $\ds{4}$ and $\ds{5}$, defined in the style
of the AdS/CFT correspondence~\cite{adsduality, adsduality2, hull,
hullkhuri, bhmds, wittds, andyds}, the masses that we have computed
translate into the energies of a dual Euclidean conformal field
theory.  Generically such theories have entropies that increase with
energy.  Therefore, if they are also to reproduce the decreasing
entropy of larger black holes, these spaces should map into ensembles
of decreasing energy.  Our results (\ref{4dmass}, \ref{5dmass}) have
precisely this property. In this regard, the fact that the 4d masses
are negative need not be worrisome because there can easily be a shift
in the formula relating energies to entropy.

It is also interesting that the numerical mass of de Sitter space in
all three examples that we have studied is the same (up to a sign) as
the numerical mass of AdS spaces in the same
dimension~\cite{stressvp}.  This may be an indication that if duals to
de Sitter exist, they may be related to the duals already known for
AdS spaces.  Indeed, in the Chern-Simons formulation of 2+1 gravity,
intriguing relations are known between the theories with positive and
negative cosmological constants.  Classical Euclidean $\ads{3}$
gravity and Lorentzian $\ds{3}$ are Chern-Simons theories of the same
group, but are endowed with different Hilbert space
structures~\cite{wittencs}.  (We hope to return to this in a later
publication.)

The total entropy of black hole and cosmological horizons
can be computed in gravity for general $m$.
It would be interesting to
test whether a Cardy-Verlinde like expression for the asymptotic
density of states of a higher dimensional CFT \cite{erik}
could reproduce this entropy.

\subsection{A bound on de Sitter masses}

Bousso has shown that, under suitable positive energy conditions on
matter fields, the entropy of de Sitter space is an upper bound on the
total entropy that can be stored in a space with a positive
cosmological constant~\cite{boussobound, raphael1}.  Above we have 
shown that our measure
of mass increases monotonically as the entropy increases, and argued
that it would be natural to map this quantity into the energy of
states in a dual field theory.  The entropy of de Sitter is then
understood as the degeneracy of such states.  Now, consider a space
which is asymptotically de Sitter in the past, but which has a mass 
measured
by (\ref{massdef}) exceeding the mass of de Sitter.  From a field
theory perspective an ensemble with this larger energy would have a
larger entropy.  Therefore, in view of the de Sitter entropy bound, we
conjecture that: 
\begin{center}
\begin{minipage}[c]{0.85\textwidth}
{\it Any asymptotically de Sitter space whose mass as defined in
(\ref{massdef}) exceeds that of de Sitter contains a cosmological
singularity.  }
\end{minipage}
\end{center}
Note that spaces with masses less than or equal to de Sitter may still
be singular for other reasons.\footnote{We have mentioned several times 
that we could have chosen to define a de Sitter stress tensor as 
$(-2/\sqrt{h}) \delta I /\delta h_{\mu\nu}$, thereby reversing the 
sign of our mass formula.  With this definition, we would be making a 
positive mass conjecture: all non-singular asymptotically de Sitter 
spacetimes have mass greater than de Sitter.}

A potential counterexample to our conjecture is provided by the
negative-mass Schwarzschild-de Sitter spacetime, in which the timelike
singularity always remains within a single cosmological region. 
However, this space is nakedly singular and the Cauchy problem is
actually not well defined -- hence it is unclear whether the space can
be admitted into consideration.  In addition, small fluctuations are
likely to lead to evolution in which the singularity propagates along
the null cone of the past cosmological horizon cutting off the ``lower
triangular'' region outside the cosmological horizon in a Penrose
diagram of de Sitter space.\footnote{We thank Rob Myers for
discussions of this point.}

\section{ RG flow vs. cosmological evolution}

Finally, we would like to consider the meaning of the holographic
\cite{holog} $UV/IR$ relation \cite{hologbound} in de Sitter space. 
Our discussion is motivated by a possible dS/CFT correspondence
\cite{hull, bhmds, wittds, andyds} in the manner of the AdS/CFT
duality~\cite{wittds,andyds}.  In analogy with the AdS/CFT
correspondence the prescription for the computation of the boundary
stress tensor, as presented above, leads quite naturally to the
relation between the trace of the stress tensor and the RG equation of
a putative dual field theory.  In particular, the precise relation
between the generators of dilatations on the boundary and the
generator of time translations in the bulk, as discussed in section
4.2, points to a natural relation between the RG flows of a possible
dual boundary theory and the time evolution in the bulk of de Sitter
space.  This statement is completely analogous to the known relation
between the RG transformations and bulk equations of motion in the
context of the AdS/CFT correspondence \cite{holorg, dvv, holorg1}.

In particular, consider the wick rotated ``kink'' solutions
of~\cite{fgpw} which were found in the context of the AdS/CFT duality. 
These solutions describe spaces interpolating between two
asymptotically $\ads{5}$ spaces and are dual to well understood RG
flows of supersymmetric 4d field theories.  Wick rotating these
``kink'' spaces produces solutions that correspond to FRW cosmologies
interpolating between two de Sitter vacua.  It is well known that RG
flows discussed by \cite{fgpw} are characterized by a positive and
monotonic $c$-function provided that a weak energy condition is
satisfied in the bulk gravitational theory.  This holographic
$c$-function coincides with the trace anomaly of the dual field
theory, and thus can be precisely related to the trace of the boundary
stress tensor~\cite{holorg1}.  In our situation, we can analogously
postulate a holographic $c$-function determined by the expression for
the trace of the Brown-York tensor and dimensional
analysis~\cite{holorg1}.  The candidate holographic $c$-function is
proportional to
\begin{equation}
c \sim \frac{1}{{(A')}^{D-2}}
\label{specialcase}
\end{equation}
in the case of a wick rotated kink solution interpolating between two
D-dimensional de Sitter spaces, where the metric of the kink solution
is
\begin{equation}
ds^2 = -dt^2 +e^A \eta_{ij} dx^i dx^j.
\end{equation}
Here prime denotes the $t$ derivative.  The asymptotic form of the
wick rotated warp factor is $A = \sqrt{\Lambda} t$.  Inserting this
expression into the formula for the holographic $c$-function gives the
correct scaling of the expected number of degrees of freedom in the
asymptotic de Sitter region (essentially determined by the value of
the holographic $c$-function at the fixed point) with the cosmological
constant $\Lambda$, as implied by the Bekenstein-Hawking entropy
formula.\footnote{These observations were originally made in
collaboration with Petr Horava.}

The correspondence between the RG evolution and the bulk cosmological
time evolution offers a nice reinterpretation of the monotonicity of
our candidate holographic $c$-function from a cosmological point of
view.  In our scenario, the IR fixed point corresponds to the period
of inflation which is driven by a large effective cosmological
constant $\Lambda_{inf}$ - the number of degrees of freedom being
proportional to $1/\Lambda_{inf}$, in 4d.  On the other hand, the UV
fixed point corresponds to an asymptotically de Sitter space with a
small positive cosmological constant $\Lambda_{obs}$ (as implied by
current observational data).  This in turn, according to the relation
between the $c$-function and the cosmological constant, corresponds to
a large number of degrees of freedom proportional to
$1/\Lambda_{obs}$.  Indeed, this is consistent with a holographic
c-theorem according to which $c_{UV} > c_{IR}$.

In the AdS/CFT correspondence, radial flow of the bulk spacetime
solutions corresponded in the dual field theory to RG
flow~\cite{holorg, dvv, holorg1}.  Here we are proposing a
relationship between the RG flow of a putative holographic dual to de
Sitter and {\it time} evolution.  In a theory of gravity, the bulk
Hamiltonian is zero, and so this is really a map between a holographic
RG equation and the Hamiltonian constraint of the bulk gravitation
theory, which at the quantum mechanical level becomes the Wheeler-de
Witt equation.  (The correspondence between boundary dilatations and
bulk time translations is also an example of the space-time
uncertainty relation \cite{yoneya}.)

More precisely, following the treatment of holographic RG flows in
asymptotically AdS spaces, we fix the gauge so that the bulk metric
can be written as
\begin{equation}
ds^2 = -dt^2 + g_{ij}dx^i dx^j
\end{equation}
The Hamiltonian constraint reads
\begin{equation}
{\cal{H}} =0
\end{equation}
where in the case of 5d bulk gravity
\begin{equation}
{\cal{H}}= (\pi^{ij} \pi_{ij} - \frac{1}{3} \pi^{i}_{i} \pi^{j}_{j})
+\frac{1}{2} \pi_{I} G^{IJ}\pi_{J} + {\cal{L}}.
\end{equation}
Here$\pi_{ij}$ and $\pi_{I}$ are the canonical momenta conjugate to
$g^{ij}$ and $\phi^I$ ($\phi^I$ denotes some background test scalar
fields). ${\cal{L}}$ is a local Lagrangian density and
$G^{IJ}$ denotes the metric on the space of background scalar fields.

As in the context of the AdS/CFT duality~\cite{dvv}, the Hamiltonian
constraint can be formally rewritten as a Callan-Symanzik equation for
the dual RG flow
\begin{equation}
\frac{1}{\sqrt{g}}( \frac{1}{3} (g^{ij} 
{\frac{\delta S}{\delta g^{ij}}})^2
-{\frac{\delta S}{\delta g^{ij}}}{\frac{\delta S}{\delta g_{ij}}}
-\frac{1}{2} G^{IJ}
\frac{\delta S}{\delta \phi^{I}} \frac{\delta S}{\delta \phi^{I}})
= \sqrt{g} {\cal{L}},
\end{equation}
provided the local 5d action $S$ can be separated into a
local and a non-local piece 
\be
S(g, \phi) = S_{loc}(g, \phi) + \Gamma (g, \phi).
\ee
In that case the Hamiltonian constraint can be formally rewritten as a 
Callan-Symanzik RG equation
\be
\frac{1}{\sqrt{g}}( g^{ij} 
{\frac{\delta }{\delta g^{ij}}} - \beta^I \frac{\delta}{\delta 
\phi^I}) \Gamma = HO
\ee
where $HO$ denotes higher derivative terms.
Here the ``beta-function" is defined (in analogy with the
AdS situation) to be $\beta^I = \partial_{A} 
\phi^{I}$
where $A$ denotes the cut-off of the putative dual Euclidean theory.

In (\ref{specialcase}) we proposed a holographic c-function for
asymptotically de Sitter spaces taking a special form.  More
generally, in analogy with the AdS/CFT correspondence~\cite{holorg1},
the holographic $c$-function can be related to the extrinsic
curvature, or equivalently, to the trace of the Brown-York tensor.  In 
a five dimensional bulk space we would have 
\begin{equation}
    c \sim \frac{1}{G \theta^3}.  
\end{equation}
Here, as we have seen
\begin{equation} <T^{i}_{i}> \sim \theta.  
\end{equation}
up to some terms constructed from local intrinsic curvature invariants
of equal time surfaces.  The RG equation is given by the conformal
Ward identity for the trace of the stress tensor 
\begin{equation}
<T^{i}_{i}> \sim \frac{d \Gamma}{d A} = \beta^I \frac{\partial \Gamma}
{\partial \phi^I}.
\end{equation}
The Raychauduri equation then implies the monotonicity of the trace of the
Brown-York stress tensor 
\begin{equation} 
    \frac{d \theta}{dt} \le 0, 
\end{equation}
as long as a form of the weak positive energy condition is satisfied
by the background test scalar fields.  This in turn guarantees the
fundamental monotonicity property we would require of a holographic
$c$-function.

Finally, there might be a holographic reinterpretation of the known
gravitational instabilities of de Sitter space~\cite{ginsperry}.  One
might speculate that the splitting of a de Sitter space into two
asymptotically de Sitter spaces~\cite{ginsperry}, could be described
by adding non-linear terms in the Wheeler-de Witt equation.  It is
tempting to conjecture that some non-linear version of the
Wheeler-deWitt equation can be related to the fully non-linear
Wegner-Wilson-Polchinski~\cite{joep} non-perturbative RG equations. 
This in turn might imply an interesting revival of the wormhole
ideas~\cite{baum} in the context of a possible dS/CFT correspondence.

\section{Discussion}

In this paper we have computed the Brown-York boundary stress tensor
of asymptotically de Sitter spacetimes and used it to define a novel
notion of mass and conserved charges.  We were motivated to carry out
this procedure in order to study the prospects for a duality between
quantum gravity on de Sitter space and a Euclidean field theory
defined on the spacelike surfaces at $\CI^{\pm}$.  The quantities we
compute would be the stress tensor and charges of the dual theory, if
such a theory were defined for de Sitter in a manner analogous to the
AdS/CFT correspondence~\cite{wittds, andyds, bhmds, hull, adsduality2}.

Several interesting results have emerged.  In all dimensions, we found
that classical objects like black holes placed in a world with a
positive cosmological constant have masses less than the mass of de
Sitter space itself, and we conjecture that all non-singular
asymptotically de Sitter spacetimes have mass less than de
Sitter.\footnote{We could have chosen the opposite overall sign for
our definition of a stress tensor and mass for de Sitter in which case
we would have a conjecture that all non-singular asymptotically de
Sitter spacetimes have masses greater than or equal to de Sitter.}
These facts make it possible that the entropy of asymptotically de
Sitter spacetime could be explained from the density of states of a
dual field theory at an energy level equal to the mass that we
measure.  Indeed, our mass formula is strongly reminiscent of the
definition of energy in a Euclidean conformal field theory.  What is
more, in three dimensions we showed that naive application of a
Cardy-like formula exactly reproduced the entropies of the Kerr-dS
spacetimes, as also observed in~\cite{park,strombousso}.  However,
despite this naive agreement, there are two important issues that need
to be addressed -- there are indications of nonunitarity (e.g., the
$L_{0}$ eigenvalues are complex in general), and there is a puzzle
regarding neglect of the Casimir energy contribution to the asymptotic
Cardy formula.

Our results might be regarded as preliminary evidence that a Euclidean
CFT dual to de Sitter space could exist.  At first sight, this not
seem to jibe with the philosophy, advocated particularly by Banks,
that the finite de Sitter entropy requires a finite dimensional
Hilbert space~\cite{banks1, bf}.  However, in the picture advocated
above, the finite entropy emerges because de Sitter space has a
particular mass and only states of quantum gravity with this energy
contribute to the entropy.  This observation, and Bousso's de Sitter
entropy bound~\cite{boussobound}, lead to a conjecture: Any
asymptotically de Sitter spacetime with mass greater than that of de
Sitter develops a cosmological singularity.  In effect, even if the
Hilbert space is formally infinite dimensional, the space of initial
data giving rise to de Sitter like evolution may be finite
dimensional.\footnote{We thank Tom Banks and Raphael Bousso for
discussions on this issue.}

The emerging picture is that Euclidean conformal field theories do
contain information about asymptotic de Sitter spaces in various
sectors.  A fascinating possibility is that these Euclidean theories
are related to the Euclidean conformal field theories that appear in
the AdS/CFT duality.  Evidence in favour of this was presented
in~\cite{bhmds} where it was argued that some Euclidean de Sitter
spaces have dual descriptions as sectors of the CFTs dual to AdS
spaces.  Further evidence, from the Chern-Simons description of 3d
gravity, appeared in the work of Witten~\cite{wittencs}.  In this work,
de Sitter and anti-de Sitter gravity were related by a change of the
Hilbert space structure in the same underlying Chern-Simons theory. 
In our case, we would define a Hilbert space structure formally by
cutting open the path integral of the euclidean field theory dual to
euclidean anti-de Sitter.  The natural conjecture is that choosing a
non-canonical Hilbert space structure would yield a dual to Lorentzian
de Sitter.  It is very tempting to adopt this idea as a working
hypothesis in trying to unearth the more precise relation between de
Sitter spaces and Euclidean conformal field theories.

\vspace{0.25in}
{\leftline {\bf Acknowledgments}}

We have benefitted from conversations with many colleagues including
Tom Banks, Raphael Bousso, Robbert Dijkgraaf, Willy Fischler, Petr
Ho\v{r}ava, Esko Keski-Vakkuri, Dietmar Klemm, Clifford Johnson, Don
Marolf, Emil Martinec, Rob Myers, Miao Li, Emil Mottola, Mu-In Park,
Simon Ross, Kostas Skenderis, Andy Strominger, Erik Verlinde and
Herman Verlinde.  {\small V.B.} and {\small D.M.} are supported
respectively by the DOE grant DE-FG02-95ER40893 and DE-FG03-84ER40168. 
{\small J.dB} was also supported in part by the NSF grant PHY-9907949
at ITP, Santa Barbara.  {\small V.B.} thanks the Helsinki Institute of
Physics, the American University in Cairo and the University of
Amsterdam for hospitality during various stages of this work.  {\small
JdB} thanks ITP, Santa Barbara where some of this work was carried
out.  {\small D.M.} thanks Harvard University, the Institute for
Advance Study, Princeton, the Caltech-USC Center for Theoretical
Physics, the Institute for Theoretical Physics, Beijing, China and the
Aspen Center for Physics for hospitality during the various stages of
this work.  Our results have been presented at a variety of venues
including string theory workshops in Amsterdam, Beijing and Aspen,
during the summer of 2001.



\begin{thebibliography}{10}
\newcommand{\wwwspires}{http://www.slac.stanford.edu/spires/find/hep/www}



\bbibitem{wald} R. M. Wald, {\em General Relativity}, University of
Chicago, 1984;~~~C.W.~Misner, J.A.~Wheeler and K.S.~Thorne, {\em
  Gravitation}, W.H.~Freeman and Co., 1973;~~~S.~Weinberg, {\em
  Gravitation and Cosmology: Principles and Applications of the
  General Theory of Relativity}.

\bibitem{hawkell}
S. W. Hawking and G. F. R. Ellis,
{\em The Large Scale Structure of Space-Time}, Cambridge University
Press, 1973.



\bbibitem{brownyork}
J.~D.~Brown and J.~W.~York,
``Quasilocal energy and conserved charges derived from 
the gravitational action,''
Phys.\ Rev.\ D {\bf 47}, 1407 (1993).



\bibitem{davjen}
D.~Kastor and J.~Traschen,
``Particle production and positive energy theorems for charged black holes in De Sitter,''
Class.\ Quant.\ Grav.\  {\bf 13}, 2753 (1996)
[arXiv:gr-qc/9311025].

\bibitem{confmass}
D.~Kastor and J.~Traschen, unpublished.;~~~T.~Shiromizu, D.~Ida and T.~Torii,
``Gravitational energy, dS/CFT correspondence and cosmic no-hair,''
JHEP {\bf 0111}, 010 (2001)
[arXiv:hep-th/0109057].

\bbibitem{hull}
C.~M.~Hull,
``Timelike T-duality, de Sitter space, large N gauge theories and
topological field theory,''
JHEP{\bf 9807}, 021 (1998)
[hep-th/9806146];~~~~
C.~M.~Hull,
``Duality and the signature of space-time,''
JHEP{\bf 9811}, 017 (1998)
[hep-th/9807127];~~~~


\bbibitem{hullkhuri}
C.~M.~Hull and R.~R.~Khuri,
``Branes, times and dualities,''
Nucl.\ Phys.\ B {\bf 536}, 219 (1998)
[hep-th/9808069];~~~~
C.~M.~Hull and R.~R.~Khuri,
``Worldvolume theories, holography, duality and time,''
Nucl.\ Phys.\ B {\bf 575}, 231 (2000)
[hep-th/9911082].


\bbibitem{bhmds}
V.~Balasubramanian, P.~Horava and D.~Minic,
``Deconstructing de Sitter,''
JHEP {\bf 0105}, 043 (2001)
[hep-th/0103171].


\bbibitem{wittds} E. Witten, ``Quantum Gravity in de Sitter space'', 
hep-th/0106109.


\bbibitem{andyds}
A.~Strominger,
``The dS/CFT Correspondence,''
hep-th/0106113.



\bbibitem{adsduality}
J.~Maldacena,
``The large N limit of superconformal field theories and 
supergravity,''
Adv.\ Theor.\ Math.\ Phys.\  {\bf 2}, 231 (1998)
[hep-th/9711200];~~~~


\bbibitem{adsduality2}E. Witten, ``Anti de Sitter Space and 
Holography'', Adv.\ Theor.\ Math.\ Phys. {\bf2}(1998) 253, 
hep-th/9802150; S.S. Gubser, Igor R. Klebanov, Alexander M. Polyakov, 
``Gauge Theory correlators from Noncritical String Theory'',
Phys.\ Lett.\ {\bf B428} (1998) 105, hep-th/9802109.
 

\bbibitem{park}
M.I.~Park, ``Statistical Entropy of Three-dimensional Kerr-De Sitter 
Space'', Phys.\ Lett.\ {\bf B440} (1998) 275, hep-th/9806119.


\bbibitem{strombousso} R.~Bousso, A.~Maloney and A.~Strominger,
``Conformal Vacua and Entropy in de Sitter Space,''
arXiv:hep-th/0112218;~~~ Talk by A.~Strominger at the ITP, Santa
Barbara and private communication from A.~Strominger and R. Bousso.


\bbibitem{klemm} 
D.~Klemm, ``Some Aspects of the de Sitter/CFT 
Correspondence'', hep-th/0106247;~~~
S.~Cacciatori and D.~Klemm,
``The asymptotic dynamics of de Sitter gravity in three dimensions,''
arXiv:hep-th/0110031.


\bbibitem{carlip}
S.~Carlip,
``The Statistical mechanics of the (2+1)-dimensional black hole,''
Phys.\ Rev.\ D {\bf 51}, 632 (1995) [gr-qc/9409052]; F. Lin and Y. 
Wu, ``Near-horizon Virasoro
symmetry and the entropy of de Sitter space in any dimension'', Phys. 
\ Rev. {\bf D453}, (1999)
222, hep-th/9901147.

\bbibitem{banados} M. Banados, T. Brotz and M. E. Ortiz, ``Quantum 
three-dimensional de Sitter space'', Phys. \ Rev. {bf D59}, (1999) 
046002, hep-th/9807216.

\bbibitem{juanandy1}
J.~Maldacena and A.~Strominger,
``Statistical entropy of de Sitter space,''
JHEP{\bf 9802}, 014 (1998)
[gr-qc/9801096].



\bbibitem{ent2}
S.~Hawking, J.~Maldacena and A.~Strominger,
``DeSitter entropy, quantum entanglement and AdS/CFT,''
hep-th/0002145.






\bbibitem{boussobound}
R.~Bousso,
``Positive vacuum energy and the N-bound,''
JHEP {\bf 0011}, 038 (2000)
[arXiv:hep-th/0010252].




\bbibitem{nariai}
H.~Nariai, ``On some static solutions of Einstein's gravitational
field equations in a spherically symmetric case,'' Sci.\ Rep.\
Tohoku Univ.\ {\bf 34}, 160 (1950);~~~~~H.~Nariai, ``On a new
cosmological solution of Einstein's field equations of gravitation,''
{\it ibid.} {\bf 35}, 62 (1951)




\bbibitem{andyrg} A. Strominger, ``Inflation and the dS/CFT 
Correspondence'',
hep-th/0110087.





\bbibitem{fgpw} D. Freedman, S. S. Gubser, K. Pilch and N. Warner,''
Renormalization group Flows from holography - Supersymmetry and a C 
theorem'',
Adv. Theor. Math. Phys. 3:363 (1999)
hep-th/9904017; L. Girardello, M. Petrini, M. Porrati and A. 
Zaffaroni,
`` Novel local CFT and exact results on perturbations of N=4
Super Yang-Mills from AdS dynamics'',
JHEP {\bf 9812}(1998) 022, hep-th/9810126.



\bbibitem{holorg}
E.~Alvarez, C.~Gomez,
``Geometric Holography, the Renormalization Group and the c-Theorem'',
Nucl.\ Phys. \ B {\bf 541} 441 (1999);~~~
V.~Balasubramanian, P.~Kraus,
``Spacetime and the Holographic Renormalization Group''
Phys.\ Rev. \ Lett. {\bf 83} 3605 (1999);


\bbibitem{dvv}
J. de Boer, E. Verlinde, H. Verlinde,
``On the Holographic Renormalization Group'',
JHEP {\bf 0008} 003 (2000).

\bbibitem{holorg1}
V.~Balasubramanian, E.~Gimon, D.~Minic,
``Consistency Conditions for Holographic Duality'',
JHEP {\bf 0005} 014 (2000);~~~
V.~Balasubramanian, E.~Gimon, D.~Minic, J.~Rahmfeld,
``Four Dimensional Conformal Supergravity from AdS Space'',
hep-th/0007211;~~~~~~
V.~Sahakian,
``Holography, a covariant c-function and the geometry of the  
renormalization group,'' 
Phys.\ Rev.\ D {\bf 62}, 126011 (2000)
[arXiv:hep-th/9910099].



\bbibitem{otherdscft}Miao Li, ``Matrix Model for De Sitter'', 
hep-th/0106184;~~~
S.~Nojiri and S.~D.~Odintsov,
``Conformal anomaly from dS/CFT correspondence,''
Phys.\ Lett.\ B {\bf 519}, 145 (2001); 
L.~Dolan, C.R.~Nappi, E.~Witten, 
``Conformal Operators for Partially Massless States'', 
hep-th/0109096; Renata Kallosh, ``N=2 Supersymmetry and de Sitter 
space'',
hep-th/0109168; C.M. Hull, ``De Sitter Space in Supergravity and M 
Theory'',
hep-th/0109213; `` Domain Walls and de Sitter Solutions of Gauge 
Supergravity'',
hep-th/0110048; B. McInnes, ``Exploring the similarities of the 
dS/CFT and AdS/CFT correspondences", hep-th/0110062; R. Kallosh, A. 
Linde, S. Prokushkin and M. Shmakova,
``Gauged Supergravities, de Sitter Space and Cosmology'', 
hep-th/0110089.


\bbibitem{abbottdeser}
L.~F.~Abbott and S.~Deser,
``Stability Of Gravity With A Cosmological Constant,''
Nucl.\ Phys.\ B {\bf 195}, 76 (1982).



\bbibitem{hensken}
M. Henningson, K. Skenderis, `` The Holographic Weyl Anomaly'', JHEP 
{\bf 9807}(1998) 023, hep-th/9806087. 



\bbibitem{stressvp}
V.~Balasubramanian and P.~Kraus,
``A stress tensor for anti-de Sitter gravity,''
Commun.\ Math.\ Phys.\  {\bf 208}, 413 (1999)
[hep-th/9902121].


\bibitem{emparan}
R.~Emparan, C.~V.~Johnson and R.~C.~Myers,
``Surface terms as counterterms in the AdS/CFT correspondence,''
Phys.\ Rev.\ D {\bf 60}, 104001 (1999)
[arXiv:hep-th/9903238].




\bibitem{haro}
S.~de Haro, S.~N.~Solodukhin and K.~Skenderis,
``Holographic reconstruction of spacetime and renormalization in the  AdS/CFT correspondence,''
Commun.\ Math.\ Phys.\  {\bf 217}, 595 (2001)
[arXiv:hep-th/0002230];~~~
K.~Skenderis,
``Asymptotically anti-de Sitter spacetimes and their stress energy  tensor,''
Int.\ J.\ Mod.\ Phys.\ A {\bf 16}, 740 (2001)
[arXiv:hep-th/0010138].





\bbibitem{mott}
P.O.~Mazur, E.~Mottola, `` Weyl Cohomology and the effective
action for conformal anomalies'', hep-th/0106151 




\bbibitem{fefgrah}
C.~Fefferman and C.R.~Graham, ``Conformal Invariants'', in
Asterisque 1985, 95.




\bbibitem{pfr}
P.~Kraus, F.~Larsen, R.~Siebelink, ``The gravitational action in 
asymptotically AdS and flat space-times'', Nucl.\ Phys.\ {\bf 
B563}(1999) 259, hep-th/9906127.





\bbibitem{deserjackiw}
S.~Deser and R.~Jackiw,
``Three-Dimensional Cosmological Gravity: Dynamics Of Constant 
Curvature,''
Annals Phys.\  {\bf 153}, 405 (1984).






\bbibitem{btz}
M.~Banados, C.~Teitelboim, J.~Zanelli, ``The black hole 
in three-dimensional space-time'', Phys.\ Rev.\ Lett. {\bf 69} (1992) 
1849, hep-th/9204099. 



\bbibitem{bh}
J.~D.~Brown and M.~Henneaux,
``Central Charges In The Canonical Realization Of Asymptotic 
Symmetries: An Example From Three-Dimensional Gravity,''
Commun.\ Math.\ Phys.\  {\bf 104}, 207 (1986).









\bbibitem{bekhaw} J. D. Bekenstein, ``Black Holes and Entropy'', 
Phys. \ Rev.\ 
{\bf D7} (1973) 2333;
S. W. Hawking, ``Particle creation by black holes'', Comm. \ Math. \ 
Phys,
{\bf 43} (1975) 199.



\bbibitem{gh1}
G.~W.~Gibbons and S.~W.~Hawking,
``Cosmological Event Horizons, Thermodynamics, And Particle 
Creation,''
Phys.\ Rev.\ D {\bf 15}, 2738 (1977).






\bbibitem{gh2}
G.~W.~Gibbons and S.~W.~Hawking,
``Action Integrals And Partition Functions In Quantum Gravity,''
Phys.\ Rev.\ D {\bf 15}, 2752 (1977).


\bbibitem{andybtz}A. Strominger, ``Black hole entropy from near 
horizon microstates'', JHEP {\bf 9802}(1998) 009, hep-th/9712251.


\bbibitem{nappi}
R.~Figari, R.~Hoegh-Krohn and C.~R.~Nappi,
``Interacting Relativistic Boson Fields In The De Sitter Universe With
Two Space-Time Dimensions,''  Commun.\ Math.\ Phys.\ {\bf 44}, 265
(1975).

\bbibitem{banks1}
T.~Banks,
``Cosmological breaking of supersymmetry or little Lambda goes back to
the future II,''
hep-th/0007146.


\bbibitem{bf}
T.~Banks and W.~Fischler,
``M-theory observables for cosmological space-times,''
hep-th/0102077.


\bbibitem{stromvaf}
A.~Strominger and C.~Vafa,
``Microscopic origin of the Bekenstein-Hawking entropy,''
Phys.\ Lett.\  {\bf B379}, 99 (1996)
[hep-th/9601029].


\bbibitem{wittencs}
E.~Witten,
``Quantization Of Chern-Simons Gauge Theory With Complex Gauge Group,''
Commun.\ Math.\ Phys.\  {\bf 137}, 29 (1991).



\bbibitem{erik}
E.~Verlinde, `` On the holographic principle in a radiation 
dominated universe'',
hep-th/0008140.


\bbibitem{raphael1}
R.~Bousso,
``A Covariant Entropy Conjecture,''
JHEP{\bf 9907}, 004 (1999)
[hep-th/9905177];~~~~
R.~Bousso,
``Bekenstein bounds in de Sitter and flat space,''
hep-th/0012052.


\bbibitem{holog}
G.~'t Hooft,
``Dimensional reduction in quantum gravity,''
gr-qc/9310026;~~~~~
L.~Susskind,
``The World as a hologram,''
J.\ Math.\ Phys.\ {\bf 36}, 6377 (1995)
[hep-th/9409089].


\bbibitem{hologbound}
L.~Susskind and E.~Witten,
``The holographic bound in anti-de Sitter space,''
hep-th/9805114; A.~Peet and J.~Polchinski, ``UV/IR Relations in AdS
Dynamics'',
Phys.\ Rev. \ D {\bf 59} 065011 (1999).






\bbibitem{yoneya} 
T.~Yoneya,
``On The Interpretation Of Minimal Length In String Theories,''
Mod.\ Phys.\ Lett.\ A {\bf 4}, 1587 (1989);~~~
M.~Li and T.~Yoneya,
``D-particle dynamics and the space-time uncertainty relation,''
Phys.\ Rev.\ Lett.\  {\bf 78}, 1219 (1997)
[arXiv:hep-th/9611072];~~~
M.~Li and T.~Yoneya,
``Short-distance space-time structure and black holes in string
theory:  A short review of the present status,'' 
arXiv:hep-th/9806240;~~~
D.~Minic,
``On the space-time uncertainty principle and holography,''
Phys.\ Lett.\ B {\bf 442}, 102 (1998)
[arXiv:hep-th/9808035].





\bbibitem{ginsperry} 
P.~Ginsparg, M.J.~Perry, `` Semiclassical perdurance of de 
Sitter space",  Nucl.\ Phys.\ {\bf B222} (1983) 245. See also, for example,
R.~Bousso, ``Proliferation of de Sitter space",
Phys.\ Rev.\ {\bf D58} (1998) 083511, hep-th/9805081; 
`` Quantum global structure of de Sitter space'', Phys.\ Rev.\ {\bf 
D60} (1999) 063503, hep-th/9902183. 
 




\bbibitem{joep}
J.~Polchinski, ``Renormalization and effective lagrangians", 
Nucl.\ Phys.\ {\bf B231} (1984) 269 and references therein. 


\bbibitem{baum} 
E.~Baum,
``Zero Cosmological Constant From Minimum Action,''
Phys.\ Lett.\ B {\bf 133}, 185 (1983);~~~
S.~W.~Hawking,
``The Cosmological Constant Is Probably Zero,''
Phys.\ Lett.\ B {\bf 134}, 403 (1984);~~~
S.~R.~Coleman,
``Black Holes As Red Herrings: Topological Fluctuations And The Loss Of Quantum Coherence,''
Nucl.\ Phys.\ B {\bf 307}, 867 (1988);~~~
S.~R.~Coleman,
``Why There Is Nothing Rather Than Something: A Theory Of The Cosmological Constant,''
Nucl.\ Phys.\ B {\bf 310}, 643 (1988);~~~
S.~B.~Giddings and A.~Strominger,
``Loss Of Incoherence And Determination Of Coupling Constants In Quantum Gravity,''
Nucl.\ Phys.\ B {\bf 307}, 854 (1988);~~~
S.~B.~Giddings and A.~Strominger,
``Axion Induced Topology Change In Quantum Gravity And String Theory,''
Nucl.\ Phys.\ B {\bf 306}, 890 (1988);~~~
 T.~Banks,
``Prolegomena To A Theory Of Bifurcating Universes: A Nonlocal
Solution To The Cosmological Constant Problem Or Little Lambda Goes
Back To The Future,''
Nucl.\ Phys.\ B {\bf 309}, 493 (1988);~~~
I.~R.~Klebanov, L.~Susskind and T.~Banks,
``Wormholes And The Cosmological Constant,''
Nucl.\ Phys.\ B {\bf 317}, 665 (1989);~~~
P.~Horava and D.~Minic,
``Probable values of the cosmological constant in a holographic theory,''
Phys.\ Rev.\ Lett.\  {\bf 85}, 1610 (2000)
[arXiv:hep-th/0001145].



\end{thebibliography}
\end{document}